\documentclass[iop]{emulateapj}

\usepackage{gensymb, float, amsmath, natbib, color, soul}
\usepackage{hyperref}
\hypersetup{colorlinks=true, linkcolor=blue, citecolor=blue, urlcolor=blue}

\slugcomment{Accepted to ApJ.}

\shorttitle{Hydrodynamic Torques in Circumbinary Accretion Disks}
\shortauthors{Moody et al.}

\begin{document}


\title{Hydrodynamic Torques in Circumbinary Accretion Disks}


\author{Mackenzie S. L. Moody, Ji-Ming Shi, and James M. Stone}
\affil{Department of Astrophysical Sciences, Princeton University, 4 Ivy Lane, Princeton, NJ 08544}
\email{mm46@princeton.edu}

%% affiliation information with \altaffiltext, with one command perpdf each


\begin{abstract}
Gaseous disks have been proposed as a mechanism for facilitating mergers of binary black holes. We explore circumbinary disk systems to determine the evolution of the central binary. To do so, we perform 3D, hydrodynamic, locally isothermal simulations of circumbinary disks on a Cartesian grid. We focus on binaries of equal mass ratios on fixed circular orbits. To investigate the orbital evolution of the binary, we examine the various torques exerted on the system. For the case where the disk plane and binary orbital plane are aligned, we find that the total torque is positive so that the semi-major axis of the binary increases. For the misaligned case, we run simulations with the binary orbital plane and disk midplane misaligned by 45\degree and find the same results - the binary grows. The timescale for the circumbinary disk to realign to the plane of the binary is consistent with the global viscous timescale of the disk.
\end{abstract}


%% subject header.  Objects should be in the appropriate "individual"
%% headers (e.g. quasars: individual, stars: individual, etc.) with the
%% additional provision that the total number of headers, including each
%% individual object, not exceed six.  The \objectname{} macro, and its
%% alias \object{}, is used to mark each object.  The macro takes the object
%% name as its primary argument.  This name will appear in the paper
%% and serve as the link's anchor in the electronic edition if the name
%% is recognized by the data centers.  The macro also takes an optional
%% argument in parentheses in cases where the data center identification
%% differs from what is to be printed in the paper.



%==================================================
%==================================================
\section{Introduction}
%==================================================
%==================================================
Circumbinary accretion disks are a common occurrence in nature, in systems such as stellar binaries \citep[e.g.][]{dutrey94, mathieu97, orosz12} and supermassive black hole binaries \citep[e.g.][]{begelman80, milosav05, dotti12}. The interactions between the binary and the disk are of great interest for understanding the evolution of the binary, and of the system as a whole. For example, the existence of circumbinary planets indicate the importance of understanding these systems, in order to understand the formation of such planets. Many such systems have been detected by the Kepler mission \citep{doyle11, welsh14}.

Black hole binaries are of particular interest, as the tools now exist to observe black hole - black hole mergers using LIGO \citep{abbott17}. It is an ongoing question what the mechanism is by which supermassive black hole binaries shrink to reach semi-major axes smaller than about a parsec, known as the final parsec problem \citep{begelman80}. A gaseous disk might pslay an important role in the evolution of the system, and may provide a mechanism for facilitating the shrinkage of the binary to these separations \citep{escala05}. Though this is a popular mechanism, the gas disk is not a definitive solution. The work of \citet{delvalle18} and \citet{souzalima17} highlight that AGN feedback in a gas disk causes the shrinkage rate of the binary to significantly decrease. \citet{souzalima17} also explore radiative cooling and find that it causes fragmentation of the disk, which leads to additional slowing of the binary decay rate.

To answer the question of the validity of gas disks as a shrinkage mechanism, it is necessary to understand the angular momentum transport within the circumbinary disk. If there were no accretion of matter on to the binary, then the binary would only lose angular momentum to the circumbinary disk, potentially presenting a solution to the final parsec problem. However, taking accretion into consideration complicates the situation, as accretion causes the binary to gain angular momentum. Therefore, the overall sign of the angular momentum transport is uncertain, and is key in understanding the evolution of the binary.

Additionally, we expect some fraction of circumbinary disks to be misaligned with the binary orbital plane. There is observational evidence that shows this misalignment in binary star systems such as HK Tau \citep{stapel98, jensen14}, KH 15D \citep{winn04}, and IRS 43 \citep{brinch16}. In binary black hole systems, \citet{nixon13} note that accretion events on to supermassive black hole binaries are likely to be chaotic and randomly oriented, as the separation of the binary is on a much smaller scale than the scale of the galaxy feeding the accretion events. Also, internal galactic processes can randomize angular momentum of the gas involved in these accretion events, further leading to the conclusion that we should not expect accretion to be aligned with the binary and therefore can expect some circumbinary disks to form misaligned. \citet{dunhill14} have simulated such accretion events onto supermassive black hole binaries, and found that this does indeed results in disks misaligned with the binary orbital plane.

There has been much work done on hydrodynamic simulations of binary systems with a circumbinary accretion disk. In particular, both \citet{miranda17} (hereafter MML17) and \citet{tang17} have done 2D simulations using a polar grid and examine the torques on the system. MML17 use the grid based code \textsc{PLUTO}, excise a central region and do not resolve the binary, and examine a range of mass ratios. They find that for circular binaries the binary semi-major axis will expand, though this changes with viscosity and binary eccentricity. \citet{tang17} perform similar simulations using the moving mesh code \textsc{DISCO}, though they resolve the whole simulation domain, and focus on the case of an equal mass binary. They examine a range of removal rates for the sink cells at the center and find that for faster sink times the net torque is positive, and for slower sink times the net torque is negative which drives the binary towards a merger. Their work builds on the work of \citet{farris14} who use the same moving mesh code and examine a range of mass ratios, but do not examine the torques on the system. In their simulations they find strong periodicity of the accretion rate which is associated with a high density lump, and for unequal mass ratios find that accretion causes the mass ratio to increase.
The work of \citet{nixon13} uses the smooth particle hydrodynamics code \textsc{PHANTOM} to perform simulations of misaligned disks for equal mass binaries, focusing on examining the tearing of the disk. For more inclined disks, they find significant tearing of the entire disk and enhanced accretion rates. \citet{fragner10} use the grid based code \textsc{NIRVANA} to perform simulations of a binary system with a slightly different configuration - there is a central primary with a secondary that is far from the disk, and misaligned from the disk midplane. They examine the disk inclination and precession evolution closely, and find that thick disks with low viscosity align with the binary orbit on the viscous evolution timescale. Thin disks with higher viscosity also align with the binary orbit on similar timescales but develop significant twists.
Most recently, the work of \citet{munoz18} builds on the work done previously in MML17. They use the moving mesh code \textsc{AREPO} to examine equal mass binaries with circular and eccentric orbits, performing a close study of the angular momentum transfer to the binary and the resulting binary orbital evolution. They find in all of their simulations that the binary expands, similar to the results of MML17.

In this paper we present 2D and 3D viscous hydrodynamic simulations of circumbinary disk systems, using the finite-volume MHD code \textsc{athena++}. We adopt a simple prescription for the disk, using a thin ($h\equiv H/R=0.1$) $\alpha$ disk model with no magnetic fields or radiative transport. The system is evolved for 1000 binary orbits, to relax to a quasi-steady state. Our main goal is to examine the evolution of the binary in systems where the binary is aligned with the disk plane, and systems where the binary and disk are misaligned. To do so, we examine the different contributions to the torque, and how this affects the long-term evolution of the binary. Our study is the first 3D hydrodynamic grid-based calculation of the entire circumbinary disk system, including the binary. \citet{shi12} have performed 3D MHD simulations of circumbinary disks but have excised the central region which includes the binary. As mentioned earlier others have previously studied misaligned disks but ours is the first study of the torques in such systems. We find that as in some previous studies, in the 2D case the binary gains angular momentum and the semi-major axis expands. More interestingly, in the both the aligned and misaligned 3D simulations the same is true - the binary grows, contrary to the common idea that the binary always loses angular momentum to the disk.

The paper is organized as follows. In Section 2, we summarize our numerical methods and analysis procedures. In Section 3, we present our results on disk properties, including the accretion rate and torques, for both aligned and misaligned disks. Finally, in Section 4 we summarize our findings and discuss the physical implications of the results.

%==================================================
%==================================================
\section{Numerical Methods}
%==================================================
%==================================================

%---------------------------------------
\subsection{Numerical Setup} \label{sec:numsetup}
%---------------------------------------
We use \textsc{athena++} \citep{stone08} as the numerical tool to solve the equations of viscous hydrodynamics: 

\begin{align}
%\begin{eqnarray}
% Continuity
  \frac{\partial \rho}{\partial t} + \nabla\cdot (\rho \mathbf{v}) = 0 \,, 
  \label{eq:continuity} \\
% Equation of motion
  \frac{\partial \rho\mathbf{v}}{\partial t} + \nabla\cdot\left(\rho\mathbf{v}\mathbf{v}
  +P\mathbf{I} + \mathbf{T_{\rm vis}} \right) = -\rho\nabla\phi \,,
  \label{eq:eom}
%\end{eqnarray}
\end{align}
where $\rho$ is the density, $\mathbf{v}$ the velocity, and $P\equiv \rho c_{\rm s}^2$ the pressure. We enforce a local isothermal equation of state \footnote{We achieve the local isothermal with adiabatic equation of state of $\gamma=1.001$, together with instantaneous cooling enforced at each timestep. } of the following form \citep{farris14}
\begin{equation}
    c_{\rm s}^2 = h^2|\phi| \,,
\label{eq:eos}
\end{equation}
where the disk aspect ratio $h\equiv H/R=0.1$ is fixed in this study. The binary potential is
\begin{equation}
    \phi = -\frac{GM_1}{|\mathbf{r} -\mathbf{r_1}|} - \frac{GM_2}{|\mathbf{r}-\mathbf{r_2}|}
\label{eq:potential}
\end{equation}
The total binary mass is $GM=GM_1+GM_2=1$, where subscript `1' and `2' denote the primary and the secondary respectively. The location of the binary members are $\mathbf{r_1}=(x_1,y_1,z_1)$ and $\mathbf{r_2}=(x_2,y_2,z_2)$ in Cartesian coordinates. For disk locations far from the binary $c_{\rm s} \propto R^{-1/2}$ where $R=\sqrt{x^2+y^2}$ is the cylindrical radius, while in the vicinity of the individual holes $c_{\rm s} \propto |\mathbf{r}-\mathbf{r_i}|^{-1/2}$ where $r=(x,y,z)$ denotes the spherical radius. In our 2D runs, the disk density $\rho$ is replaced by the surface density $\Sigma$, and the pressure is
therefore the vertically integrated value. 

In this work, we choose isotropic viscosity to account for the angular momentum transport within the disk. $\mathbf{T}_{\rm vis}$ is the viscous stress tensor, defined as 
\begin{equation}
    \mathbf{T}_{{\rm vis}, i\,\!j} = -\rho\nu\left(\frac{\partial v_i}{\partial x_j}+\frac{\partial v_j}{\partial x_i}-\frac{2}{3}\frac{\partial v_k}{\partial x_k}\delta_{ij} \right)
\label{eq:stress_tensor}
\end{equation}
in its component form, where $x_i, x_j, x_k \in [x,y,z]$. The isotropic viscous coefficient is defined following the $\alpha$-disk prescription \citep{shakura73} $\nu = \alpha c_s^2/\Omega$ with $\alpha=0.1$ and $\Omega=\Omega_K \equiv (GM/R^3)^{1/2}$ where $R=\sqrt{x^2+y^2}$ the cylindrical radius. MML17 have carried out simulations using both $\alpha=0.1$ and $\alpha=0.05$ and find the same result for circular binaries, so we do not vary $\alpha$ in our study. Numerically, the stress tensor is calculated explicitly and added directly to the hydro momentum flux, while the binary forcing term is added to the right hand side of equation \ref{eq:eom} as a source term. 

The initial disk density profile is 
\begin{equation}
    \rho(R,z)|_{t=0} = \rho(R,z=0) \exp\left[\frac{GM}{c_s^2}\left(\frac{1}{\sqrt{R^2+z^2}}-\frac{1}{R}\right) \right]
\label{eq:init_density}
\end{equation}
where 
\begin{equation}
    \rho(R,z=0) = \rho_0 (R/a)^{-p} \exp \left[-(R/R_{s})^{-2} \right]
\end{equation}
is the midplane density. We set density unit $\rho_0 = 1$, binary separation $a=1$ as the length unit, and the time unit $\Omega_{\rm bin} \equiv \sqrt{GM/a^3} =1 $. For our 3D simulations $p=1.5$ so that it matches $\Sigma\propto R^{-1/2}$ as in the 2D runs, where $p=0.5$. To facilitate a faster steady solution, we start with a gap interior $R\lesssim R_s=2.5a$ similar to the work of \citet{tang17}. We correct the initial azimuthal velocity $v_{\phi}$ of the disk with the pressure effects from the local pressure profile and the binary's quadruple potential term. The initial radial velocity $v_R$ takes into account of the small radial drift due to viscous accretion. 

The binary is assumed to be of equal mass, and orbits on a fixed circular orbit with separation $a=1$. In both the 2D and 3D in-plane cases, the binary and the disk share the same orbital plane. In the 3D misaligned case, the binary is out of the $z=0$ disk plane with a fixed inclination angle $i=\pi/4$.

The computational domain extends to $[-10a,10a]\times[-10a,10a]\times[-5a,5a]$ in $x$, $y$, and $z$ respectively. We choose our boundary conditions along $x=\pm 10a$ and $y=\pm 10a$ direction to be outflow, and the vertical boundary conditions to be diode so that inflow is not allowed. A damping zone between $R = R_i = 8a$ and $R_o=10a$ is imposed to quench the $m=4$ modes according to the following formula:
\begin{eqnarray}
du/dt = - [u-u(t=0)]\sl{F} \\
\sl{F} = \frac{R^2-R_{i}R}{R_o^2-R_iR_o}t_{\rm orb}(R_i)^{-1}
\end{eqnarray}
where $u$ represents the hydro variables and $t_{\rm orb} = 2\pi/\Omega_K$ is the orbital period at $R$.

We also simulate the accretion process with sink cells centered around each individual binary members with a spherical radius $r_{\rm sink} = 0.1a$. Within this radius, we remove the mass according to the $\alpha$-disk model 
\begin{equation}
    d\rho/dt = -\rho/t_{\rm rm}\,,
\end{equation}
with the mass remove timescale 
\begin{equation}
    t_{\rm rm} = t_{\rm visc}(|\mathbf{r}-\mathbf{r_i}|) = \frac{2|\mathbf{r}-\mathbf{r_i}|^2}{3\nu}
\end{equation}
This is the same sink prescription used by \citet{tang17}. Our fiducial runs make use of the static mesh refinement of \textsc{athena++}. At the root level, the resolution is $256\times256\times128$ on the aforementioned $[-10a,10a]\times[-10a,10a]\times[-5a,5a]$ computational domain. We further make 2-level refinement for the regions $[-5a,5a]\times[-5a,5a]\times[-2.5a,2.5a]$ and $[-2.5a,2.5a]\times[-2.5a,2.5a]\times[-1.25a,1.25a]$. We therefore resolve the sink cell regions with roughly $10$ cells across. 

%---------------------------------------
\subsection{Analysis Procedure} \label{sec:torquecalc}
%---------------------------------------
As in the work of MML17 and \citet{tang17}, we would like to examine the effects of the torque on the evolution of the binary. To do so, we calculate the advective, viscous, gravitational, and total torques on the system. The definitions of each are as follows:

\begin{equation}
    \mathbf{\dot{J}}_{adv} = \int \rho \mathbf{v} \times \mathbf{r} \  (\mathbf{v} \cdot \mathbf{r}) \  dS
\end{equation}
\begin{equation}
    \mathbf{\dot{J}}_{visc} = - \int (\mathbf{T}_{visc} \cdot \mathbf{\hat{e_r})} \times \mathbf{r} \ dS
\end{equation}
\begin{equation}
    \mathbf{T}_{grav} = \int \rho \ (\mathbf{r} \times \mathbf{\nabla}\phi) \ dS
\end{equation}
\begin{equation}
    \mathbf{\dot{J}}_{tot} = \mathbf{\dot{J}}_{adv} - \mathbf{\dot{J}}_{visc} - \mathbf{T}_{grav}
    \label{eq:totaltorq}
\end{equation}

The torques are calculated within the code at every timestep. Each component of the integral ($x, y, z$) is calculated separately over spherical shells throughout the disk. Each spherical shell has a width of two cells ($2\Delta x$), in order to get a smooth average. The torque data is output instantaneously every tenth of a binary period. In our calculations we use the data from the last 100 orbits in order to ensure we are analyzing the disk in a steady state.

Once calculated, these are used to examine if the binary is expanding or contracting. Though we have all three components of the torque, we consider the z-component of the torque only, as this is the contributing factor to the binary evolution. We examine the binary evolution using the following expression as described by MML17, which assumes a fixed circular orbit:

\begin{equation} \label{eq:evolution}
    \frac{\dot{a}}{a} = 8 \Big( \frac{l_0}{l_B} - \frac{3}{8}\Big) \frac{\dot{M}}{M}
\end{equation}

Here, $l_B = [GMa(1-e_B^2)]^{1/2}$ is the specific angular momentum of the binary, and $l_0$ is ${\langle \dot{J}\rangle}/{\langle \dot{M}\rangle}$ calculated from the average total torque described in equation \ref{eq:totaltorq}. The accretion rate is calculated in spherical shells, described for the torque components, and then averaged over radius. $GM$ and $a$ are 1 in our simulations and the binary is on a fixed circular orbit so $e_B =0$, giving us a value for $l_B$ of 1.  Here we are concerned with whether the binary is shrinking or growing, not the rate at which this is happening. This can be determined from the sign of $\dot{a}$, so the $\dot{M}/{M}$ term does not need to be calculated. From this, we see that the sign of $\dot{a}$ is determined by whether $l_0$ is less than or greater than $3l_B/8 = 3/8$.

In our misaligned disk simulations we calculate the inclination angle through the disk, to determine how well the disk is aligned with the binary orbital plane. We use the definition of inclination angle through the disk, $\delta$, as described by \citet{fragner10}:
\begin{equation}
    \cos{\delta} = \frac{\mathbf{J_B \cdot L}}{\mid \mathbf{J_B}\mid \mid \mathbf{L} \mid}
\end{equation}
where $\mathbf{J_B}$ is the angular momentum vector of the binary and $\mathbf{L}$ is the total angular momentum vector at some radius in the disk. As with the torques, $\mathbf{L}$ is calculated over spherical shells. We then use the inclination angle to determine the realignment time of the disk, i.e. the time it will take for the circumbinary disk to realign to the binary orbital plane. This is compared to the viscous time, which is calculated using the following:
\begin{equation}
    T_{\nu} = \frac{R^2}{\nu} = \frac{1}{\alpha h^2}\sqrt{\frac{R^3}{GM}}.
\end{equation}

%==================================================
%==================================================
\section{Results}
%==================================================
%==================================================

\begin{deluxetable}{l l l l l}
\tablecaption{Simulation Parameters \label{tab:table1}}
\tablehead{\colhead{Run} & \colhead{Dim} & \colhead{Resolution} & \colhead{Inclination} & \colhead{Sink time ($t_{\textrm{rm}}$)}}
\startdata
A & 2D & $256\times256$ & - & 29.814 \\ 
B & 2D & $256\times256$ & - &  5.0 \\
C & 2D & $256\times256$ & - & 0.1 \\
D & 2D & $128\times128$ & - & 29.814 \\
E & 3D & $256\times256\times128$ & 0\degree & 29.814 \\
F & 3D & $256\times256\times128$ & 45\degree & 29.814
\enddata
\end{deluxetable}

In table \ref{tab:table1} we show the different simulations we have run with their different properties. All simulations have the same disk aspect ratio ($h\equiv H/R = 0.1$) and $\alpha$ ($\alpha = 0.1$). The fiducial runs are runs A, E, and F. In runs B and C we test faster sink times for the sink cells representing the black holes. In run D we test a lower resolution to test convergence.

%---------------------------------------
\subsection{2D Simulations}
%---------------------------------------
First we run 2D simulations to check our numerical setup and the reliability of the Cartesian grid. Many similar simulations have been done already, as seen in the work of MML17, \citet{farris14}, \citet{tang17}, and \citet{munoz18}. None use a Cartesian grid in their simulations. The work of \citet{farris14} do not examine the torques on the system though the other three do. In addition, the work of MML17 excise a central region so they do not resolve the binary itself.

\subsubsection{General Properties}

\begin{figure}
 \centering
 \includegraphics[width= \linewidth]{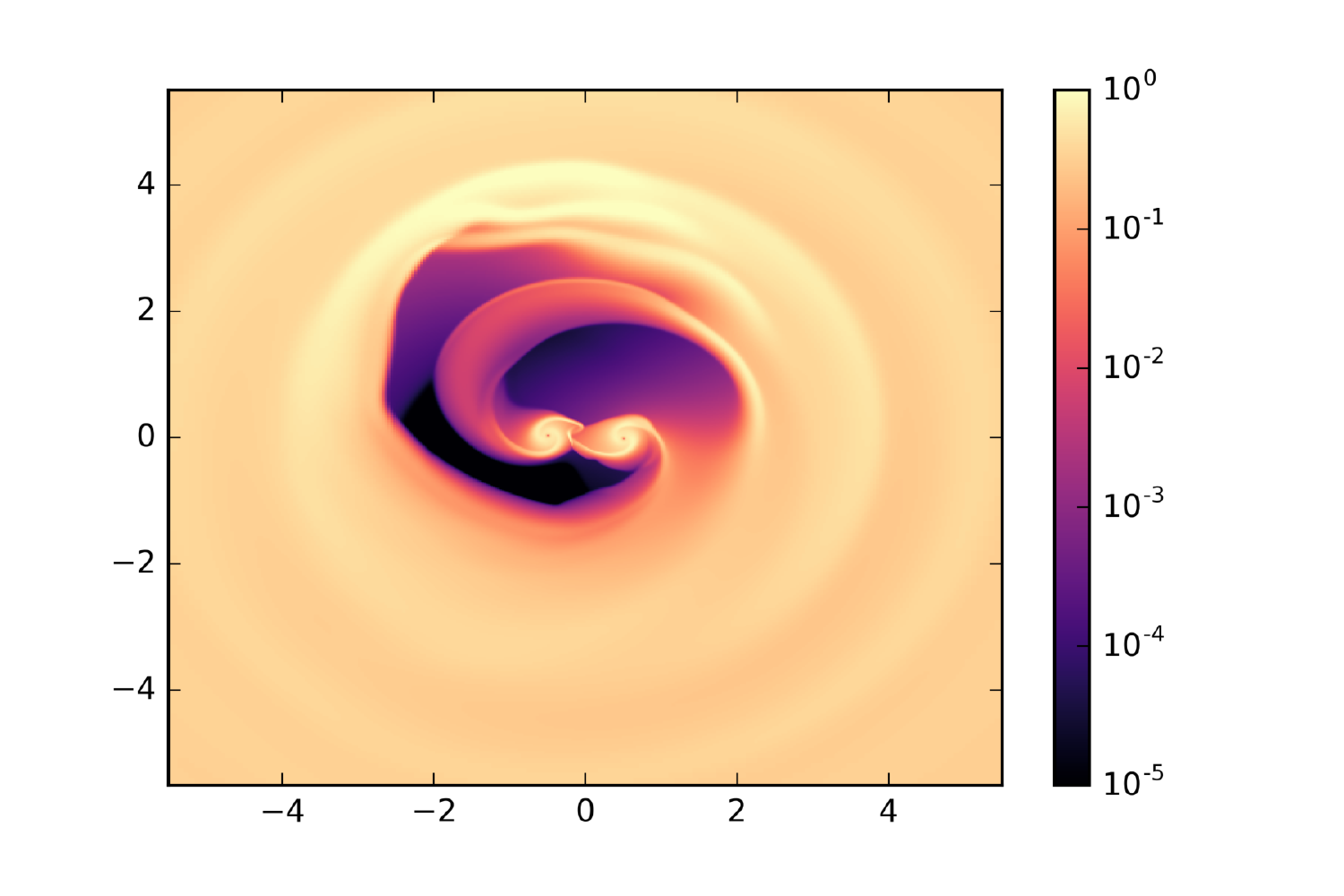}
   \caption{Snapshot of the surface density $\Sigma$ at $t \approx 1000$ binary orbits for our 2D simulation with resolution of $256\times256$. Density is shown in a logarithmic scale indicated by the side bar, and the $x/y$ axes are in units of binary separation ($a$, set to 1). The orbital motion of both the binary and gas disk are in the counterclockwise direction.}
   \label{fig:2d256dens}
\end{figure}

\begin{figure}
 \centering
 \includegraphics[width= \linewidth]{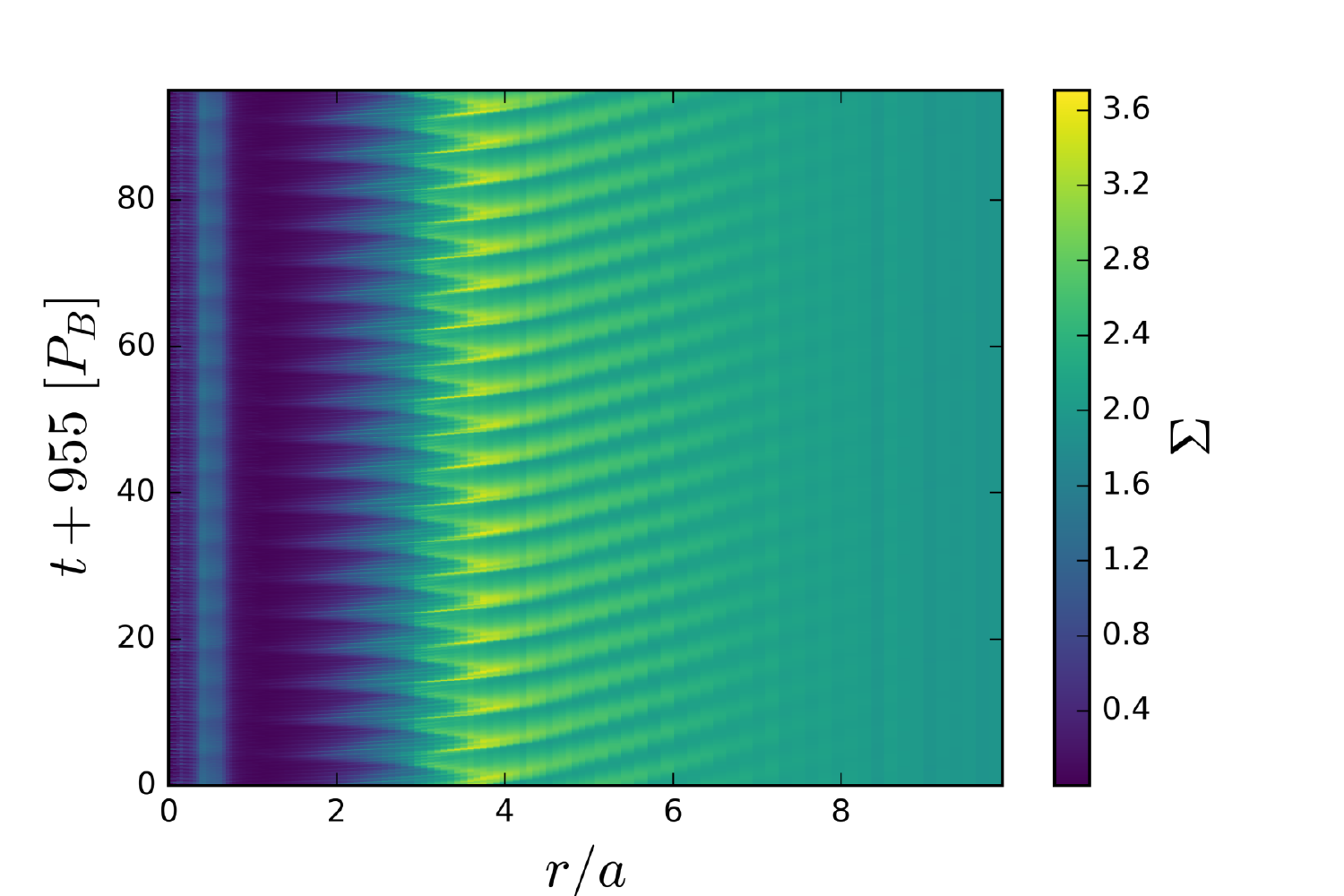}
   \caption{Spacetime diagram of surface density of the disk for the last 100 binary orbits. The $x$ axis is radius from the center given in units of binary separation ($a$, set to 1). The $y$ axis is time in binary orbits. The density is shown in the side bar, in a linear scale. The regular pattern seen here indicates spiral density waves.}
   \label{fig:2d256spacetime}
\end{figure}

Our simulation (simulation A as described in table \ref{tab:table1}) has a resolution of $256\times256$ at each level of mesh refinement, which gives us a resolution of about 102 cells per $a$ in the innermost region of $[-2.5a,2.5a]\times[-2.5a,2.5a]$. The viscous time at $r=a$ is 119 binary orbits, and the global viscous time of the disk is 5020 binary orbits. We run the simulation for 1050 binary orbits, about 1/5 of the global viscous time of the disk, in order to reach a steady state. This requires approximately 140 CPU hours to complete the simulation. 

A snapshot of the surface density of the system at late time is shown in Figure \ref{fig:2d256dens}. Only the inner part of the disk is shown. We can see clearly the circumbinary disk with a gap region cleared out around the central binary as expected, with a gap radius of roughly $r=2a-3a$. MML17 have the same size of the central gap region in their simulation with the same parameters as we have used. The gap is generally not centered about the central binary in both our simulation and those of MML17. Additionally, we see towards the top of the cavity in our simulation a further asymmetry in the gap region. A high density lump has formed in the disk at the edge of the gap. This lump is created from the material that is flung back towards the disk away from the binary. The lump orbits along with the disk, and is periodically torn apart and recreated as it orbits. 

Around each black hole, we can also see the formation of mini disks, with accretion streams feeding the mini disks from the circumbinary disk. MML17 see the same features in their simulations (spiral density waves, accretion streams, and lump), though they are unable to see mini disks as they have excised the central region. The simulations of \citet{tang17} have the same disk aspect ratio and $\alpha$ values as our simulation, though they use a polar grid and a slightly different sink time. Their simulations are also consistent with ours as they see the same size asymmetric gap region, density lump, accretion streams, and spiral density waves. \citet{tang17} do resolve the inner portion of the disk, and also see mini disks that form around each black hole with accretion streams feeding them.

To analyze if the system has indeed reached a steady state we plot the integrated surface density profile of the disk as a function of radius and time, seen in Figure \ref{fig:2d256spacetime}. This figure shows the last 100 binary orbits of the simulation. The main structure seen here is the spiral density waves, which are excited by the binary and travel outwards. The lack of secular trends is consistent with the disk having reached a steady state by the end of our simulation time. 

\subsubsection{Accretion Rate} \label{sec:2dmdot}

\begin{figure}
 \centering
 \includegraphics[width= \linewidth]{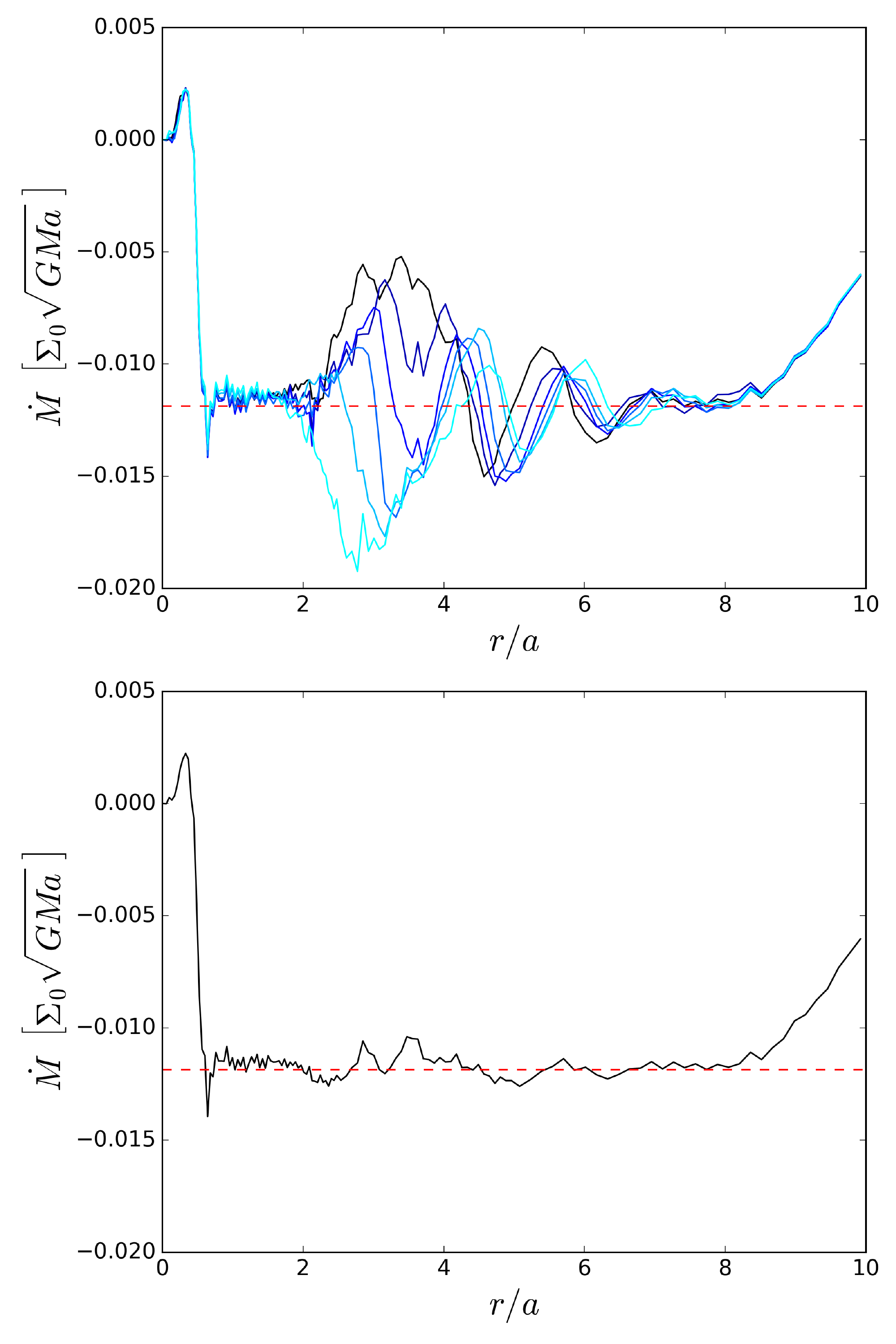}
   \caption{Accretion rate in the disk as a function of radius for the $256\times256$ simulation run. In the upper panel, we plot accretion rate for the last 60 orbits, with each line representing an average over 10 orbits. The lines follow the black to lightest blue color gradient progressively in time. The red dashed line indicates the overall average of the accretion rate, averaged over time (the last 100 orbits) and over radius (excluding the buffer zone of the disk). The lower panel shows the accretion rate through the disk, averaged over the entire period of the last 100 orbits. The red dashed line is again the overall average, averaged over radius again excluding the buffer zone. The overall average rate is 0.012 $\Sigma_0 \sqrt{GMa}$.}
   \label{fig:2d256mdot}
\end{figure}

The accretion rate is another indicator of whether we have reached a steady state. In Figure \ref{fig:2d256mdot} we show the accretion rate through the disk. The upper plot shows the accretion rate through the disk for the last 100 orbits of the simulation, with each line representing an average over 10 orbits. The red dashed line indicates the average accretion rate, averaged over time and radius for these last 100 orbits. The lower plot is an average over the entire period of the last 100 orbits. The red dashed line is the same as in the upper plot - the average accretion rate averaged over both time and radius. The average rate is 0.012 $\Sigma_0 \sqrt{GMa}$. This value is larger than for an $\alpha$-disk around a single black hole by a factor of 1.27. \citet{farris14} also find their disks to have larger accretion rates than single black hole disks. They note that this does not indicate that binaries enhance accretion, but rather that binaries are unable to fully clear the gap region and suppress accretion. 

Although there is larger oscillation on shorter timescales as shown in the upper plot of Figure \ref{fig:2d256mdot}, the lower plot indicates that the average accretion rate is generally steady for the majority of the circumbinary disk. This behavior is consistent with the simulations in MML17. The most variation occurs around about $r=3a$, as seen in more detail in the upper plot. This region is near where the edge of the gap region is located, so there is interplay between material being accreted to the mini disk and some material being flung outwards due to the torque. This variation is the same as we see in Figure \ref{fig:2d256spacetime}, which indicates this is where the inner edge of the disk oscillates in and out. We conclude that mass is flowing through the disk at a roughly steady rate.

\begin{figure}
 \centering
 \includegraphics[width= \linewidth]{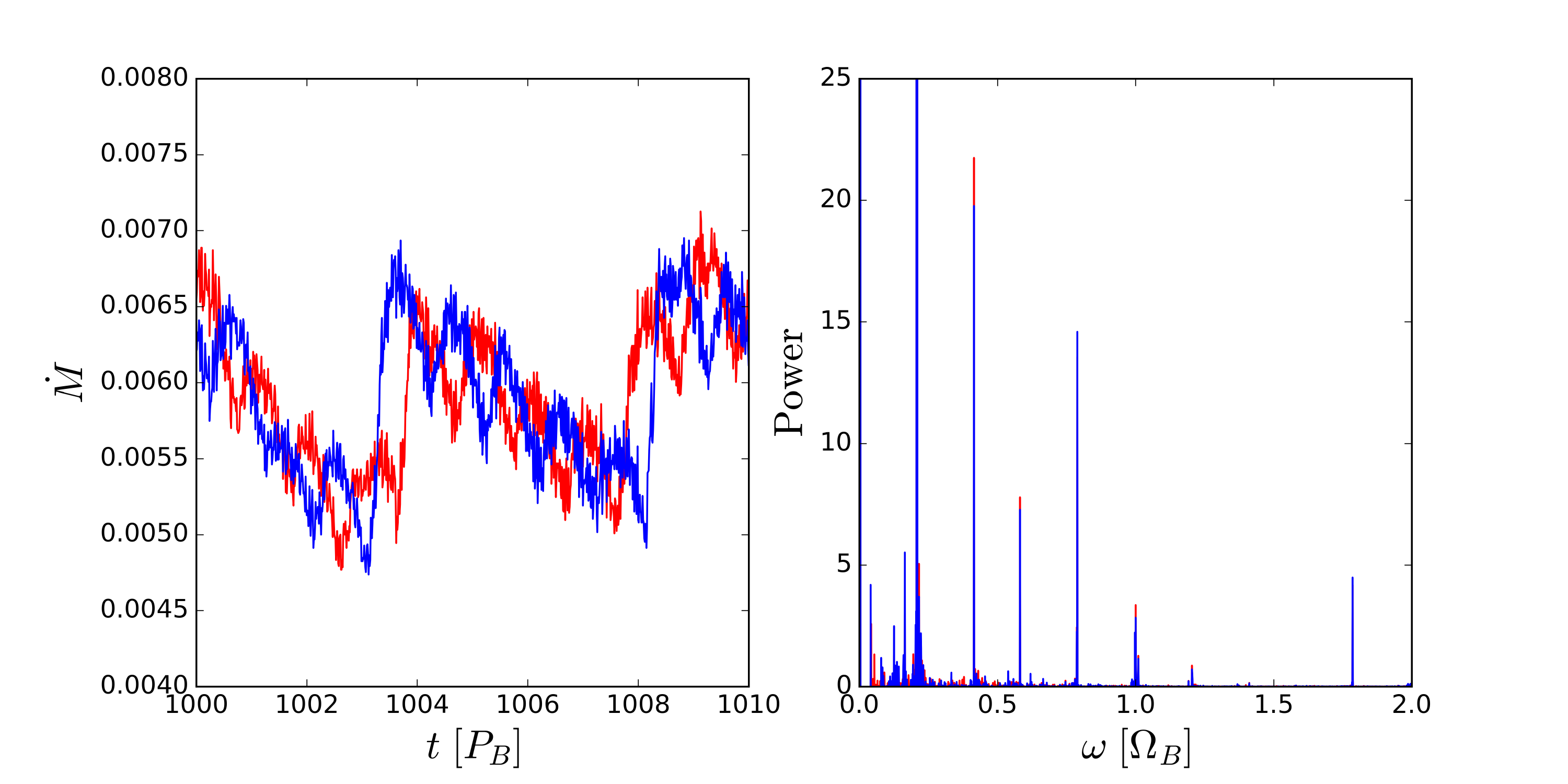}
   \caption{Accretion rate for the black holes as measured by recording the mass removed by the sink cells representing the black holes, for the 2D simulation (resolution $256\times256$). The left-hand panel shows the accretion rate for each of the individual black holes for a period of ten orbits near the end of the simulation. The right-hand panel shows the power spectrum of the accretion rates, for the entire simulation run time.}
   \label{fig:2D256BHacc}
\end{figure}

We also record the amount of mass removed over time by the sink cells representing the central black holes. In doing so, we find that a single black hole has an accretion rate of 0.00559 $\Sigma_0 \sqrt{GMa}$ giving a total accretion rate of 0.0112 $\Sigma_0 \sqrt{GMa}$. Comparing to our calculation of accretion rate through the disk we find the two calculations agree to within 7\%. The accretion rate calculated by the sink cells is shown in Figure \ref{fig:2D256BHacc}. In the left-hand panel the most obvious feature is the burst-like variability on the timescale of about 5 binary orbits. MML17 see the same variability with the same frequency. They note that their power spectrum peaks at about $\Omega_B /5$ ($\omega = 0.2$). We plot the power spectrum of the accretion rates in the right-hand panel of Figure \ref{fig:2D256BHacc}. Our largest peak occurs at 0.2075, nearly the same value. This burst-like variation is associated with presence of the high density lump at the edge of the disk gap. There are other periodic variations with higher frequencies in the left-hand panel, reflected in the many other peaks in the power spectrum. These peaks are seen by MML17 in their simulations of circular binaries. \citet{farris14} have also performed similar 2D simulations for circular binaries (though unlike MML17 they do resolve the inner cavity) and find very similar behavior in the accretion rate. Their results for an equal mass ratio also show that the dominant peak occurs around the same frequency of 0.2.

\subsubsection{Torques} \label{sec:torque2D}

\begin{figure}
 \centering
 \includegraphics[width= \linewidth]{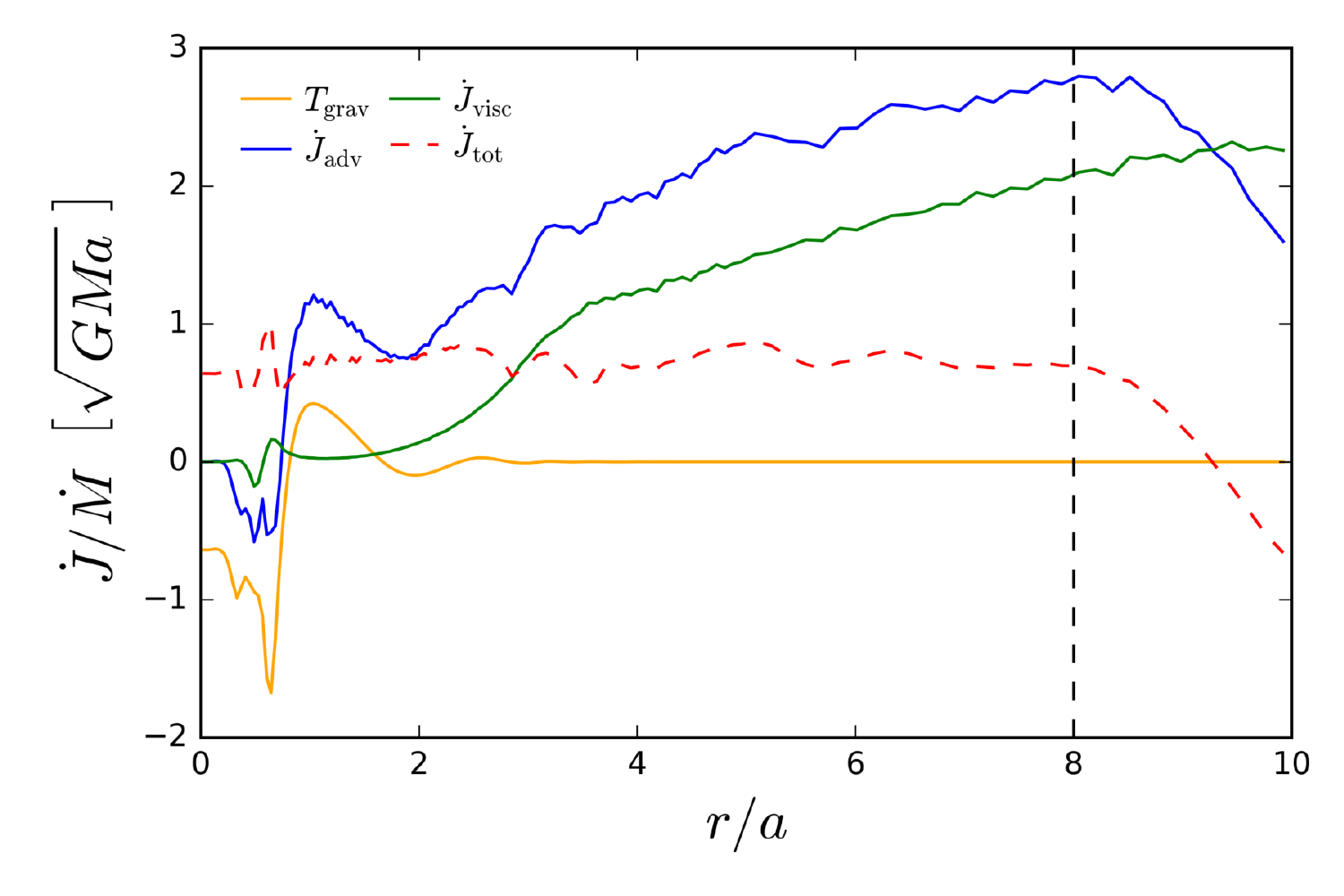}
   \caption{Advective, viscous, gravitational, and total torques on the system for the $256\times256$ 2D simulation, normalized by the average accretion rate. Torques are calculated as described in Section \ref{sec:torquecalc}.}
   \label{fig:jflux2d256}
\end{figure}

In Figure \ref{fig:jflux2d256}, we show the time-averaged advective, viscous, and gravitational torques, as well as the total net torque. All torque components are normalized by the average accretion rate, 0.0112 $\Sigma_0 \sqrt{GMa}$. Our results are very similar to what is seen in Figure 17 of MML17. The main difference occurs after $r=8a$ as our disk has a buffer zone whereas their disk extends much further out. We focus our comparison on our results outside $r=1a$ as their work has excised the central region. All torque components in our simulation have the same profile as in MML17, and the features in the profiles occur at the same values of $r$ and $\dot{J}/\dot{M}$. Our total net torque (the red dashed line in Figure \ref{fig:jflux2d256}) is remarkably flat with the largest variation at about $r=0.5a$. This is due to this radius being the location of the binary. The overall average value of the normalized net torque for our 2D simulation is 0.723  $\sqrt{GMa}$, which is close to the value of about 0.8 stated in MML17. The simulation of a circular equal mass binary in the work of \citet{munoz18} (shown in their Figure 3) shows almost identical behavior to both our simulation and that of their own previous work in MML17. The value of their normalized torque is 0.68, in good agreement with our result.

As we see the same positive net torque as MML17 and \citet{munoz18} with the same shape and value, evolution of the binary orbit will result in binary expansion. By taking the average of the total torque curve, we now have our $l_0$ value needed to utilize equation \ref{eq:evolution}. As $0.723 > 3l_B/8$ we have a positive $\dot{a}$, indicating that the binary will be expanding, the same result as found in MML17 and \citet{munoz18}.

As we have noted in Section 3.1.1 the work of \citet{tang17} is very similar to the 2D simulations we have run. They find a dependence of the torque on the sink time, with rapid sink times resulting in positive torques and slower sink times resulting in negative torques. Though we run a slow sink time, our simulation therefore seems to correspond with their rapid sink time results as we find positive torques. We have varied our sink time to explore the effect on the torques and have not found the same result as \citet{tang17}, we instead find that the torques in our simulations are not sensitive to the sink time. This is described in more detail in Section 3.1.5. The differences between our work and theirs may be related to the numerical methods used (we use a grid-based code and they use a moving-mesh code) or the initial density profile, which is slightly different. We have carried out a convergence study to make sure our results are not affected by numerical diffusion, described in the following section (Section 3.1.4).

\subsubsection{Resolution Test}

\begin{figure}
 \centering
 \includegraphics[width= \linewidth]{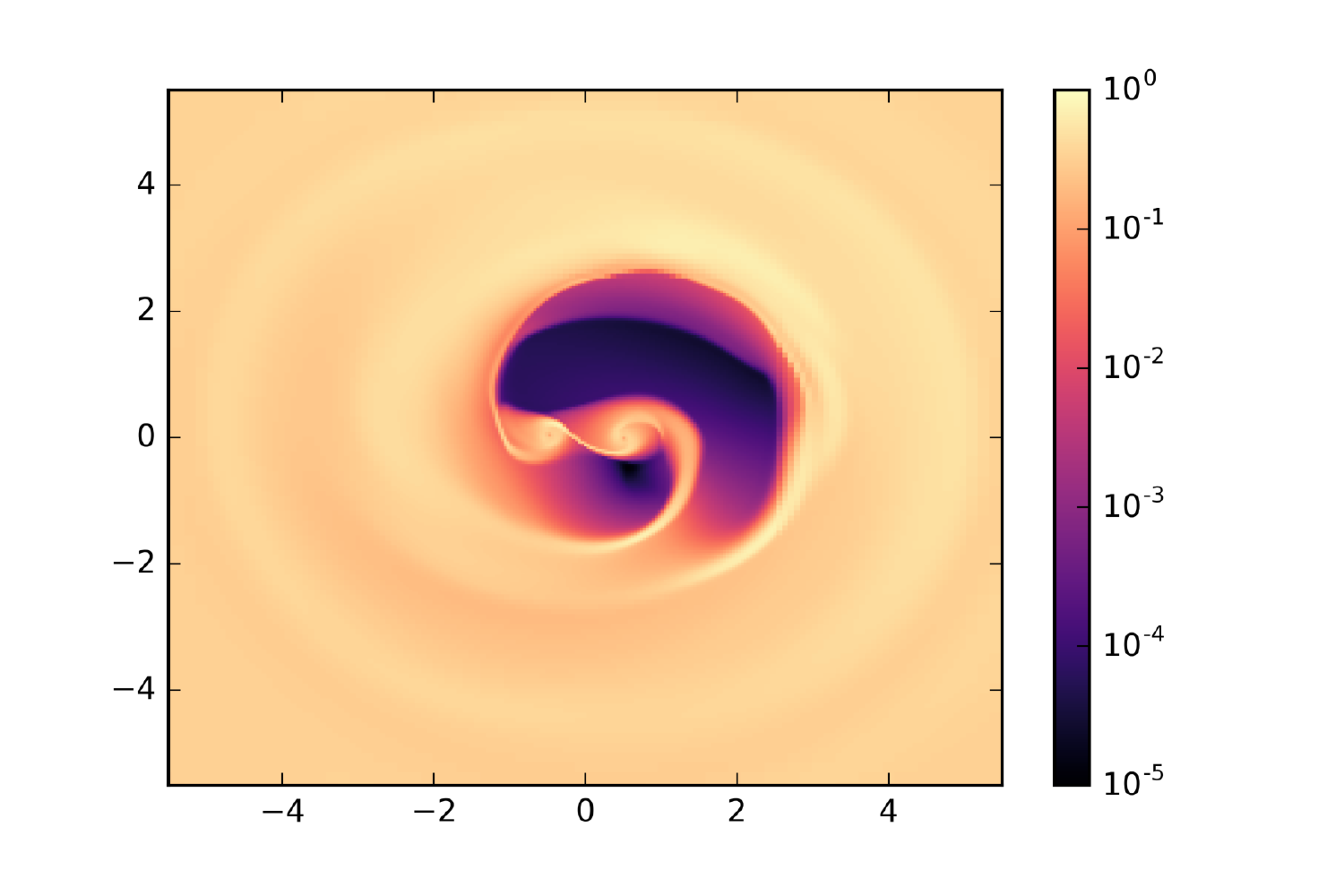}
   \caption{Snapshot of the surface density $\Sigma$ at $t \approx 1000$ binary orbits for our 2D simulation with resolution of $128\times128$. Density is shown in a logarithmic scale indicated by the side bar, and the $x/y$ axes are in units of binary separation ($a$, set to 1). The orbital motion of both the binary and gas disk are in the counterclockwise direction.}
   \label{fig:2d128dens}
\end{figure}

\begin{figure}
 \centering
 \includegraphics[width= \linewidth]{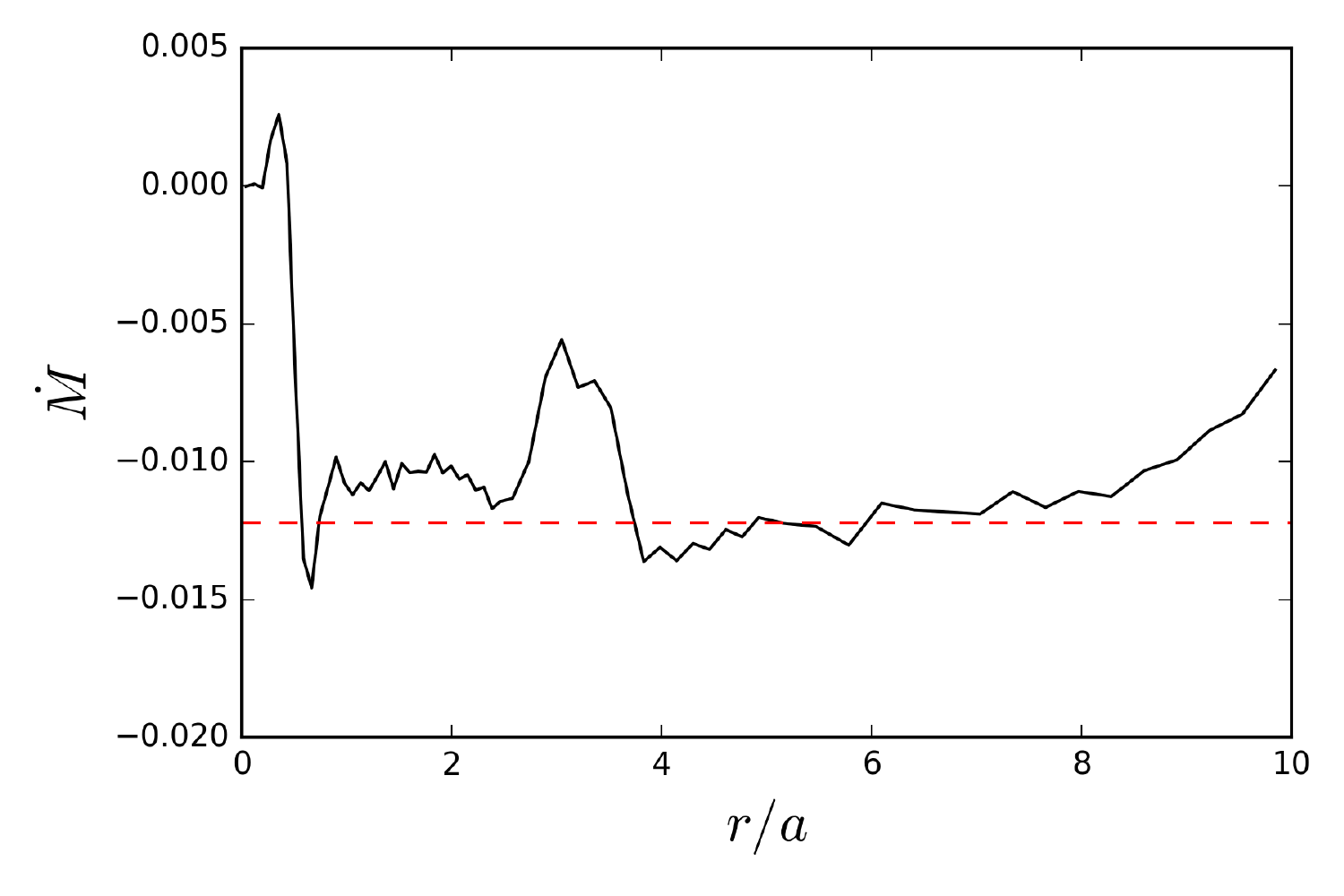}
   \caption{Accretion rate in the disk as a function of radius for the $128\times128$ simulation run. This is the accretion rate through the disk, averaged over the entire period of the last 100 orbits. The red dashed line is the overall average, averaged over radius again excluding the buffer zone. The overall average rate is 0.012 $\Sigma_0 \sqrt{GMa}$, the same as the higher resolution run.}
   \label{fig:2d128mdot}
\end{figure}

We have also performed a 2D simulation with resolution of $128\times128$, keeping all other simulation properties identical to the $256\times256$ simulation (simulation D as described in table \ref{tab:table1}). The surface density for the same time in the simulation as in Figure \ref{fig:2d256dens} is shown in Figure \ref{fig:2d128dens}. From this figure we can see the same general structure of the system: the gap region within $ r=2.5a$, spiral density waves, mini disks around the individual black holes, and accretion streams. We calculate accretion rate and normalized torque for this lower resolution simulation. The two 2D simulations show the same overall accretion rate through the disk. This is shown in Figure \ref{fig:2d128mdot}, which is the equivalent of the lower panel of Figure \ref{fig:2d256mdot} but for this lower resolution run. Though there is variation in the accretion rate over time in both simulations, the shape of the accretion rate as a function of radius is the same, with lower variation in the $256\times256$ simulation than the $128\times128$ simulation. We find the same trend when we analyze the torque in the lower resolution simulation. The torques have the same numerical value, and the curves as a function of radius have the same behavior as the higher resolution run. Again, we do see slightly larger variation in each torque component curve in the lower resolution run. We conclude that the simulation is converged based on the comparison between the two resolutions showing the same results with very minor differences, which are small variations on the same results.

\subsubsection{Comparing Sink Times}
\begin{figure}
 \centering
 \includegraphics[width= \linewidth]{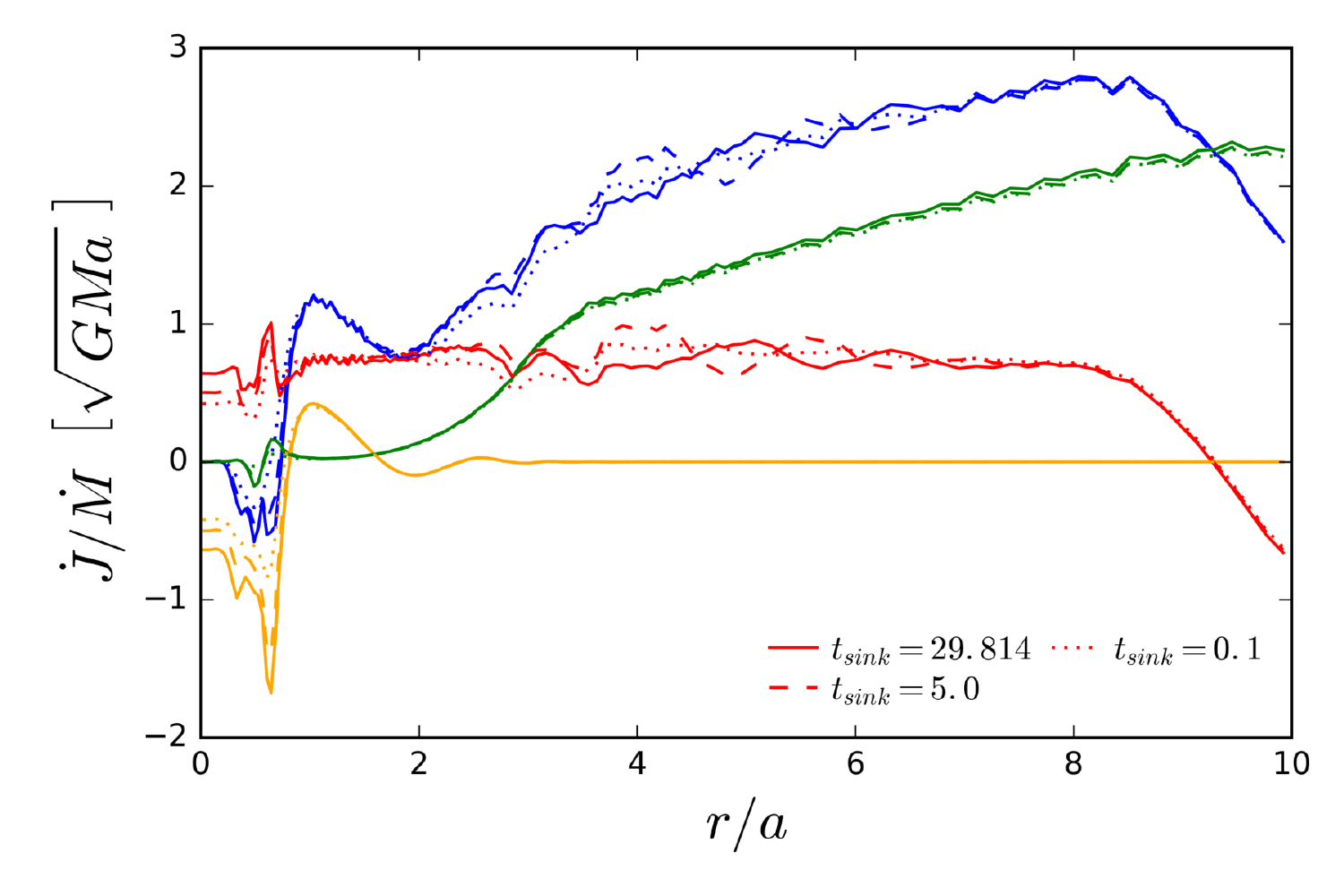}
   \caption{Advective, viscous, gravitational, and total torques on the system for $256\times256$ 2D simulations with various mass removal timescales, normalized by the average accretion rate. Torques are calculated as described in Section \ref{sec:torquecalc}. The solid lines represent $t_{rm} = 29.814$, dashed lines are $t_{rm} = 5.0$, and dotted lines are $t_{rm} = 0.1$. The different colors are the same as in Figure \ref{fig:jflux2d256} (advective is blue, viscous is green, gravitational is orange, and total net torque is red).}
   \label{fig:sinkcomp}
\end{figure}

As stated in section \ref{sec:numsetup} for this 2D simulation we have used the viscous time at the sink radius as the mass removal timescale (the sink time), which is $t_{rm} = 29.814$. We now test different sink times to see if there is a relationship between the resulting torques and the mass removal rate. In Figure \ref{fig:sinkcomp} we show the torque components and net torque for three different sink times, $t_{rm} = 29.814$, $t_{rm} = 5.0$, and$t_{rm} = 0.1$ (these are simulations A, B, and C as described in table \ref{tab:table1}). The overall behavior is the same in all the curves and the average net torque is within 4\% of the simulation with $t_{rm} = 29.814$ for both of the other sink times. As we see no discernible difference in the sink time we use the viscous time ($t_{rm} = 29.814$) for the 3D simulations that follow.

%---------------------------------------
\subsection{3D Simulations - aligned disk}
%---------------------------------------
To examine if the same effect on binary evolution is seen in 3D we first perform 3D simulations with the binary orbiting in the same plane as the circumbinary disk, to test if the assumptions made in 2D simulations are valid. Some work in 3D has been done by \citet{shi12}, though they perform full MHD simulations rather than pure hydro simulations. Our simulation is simulation E as described in table \ref{tab:table1}. As mentioned in Section \ref{sec:numsetup} our resolution is $256\times256\times128$ on a computational domain of $[-10a,10a]\times[-10a,10a]\times[-5a,5a]$ in $x$, $y$, and $z$, with two levels of mesh refinement. In order to compare with our 2D simulations, we run the simulations for the same amount of time - 1050 binary orbits, equivalent to about 1/5 of the global viscous time of the disk. This requires about 106000 CPU hours to complete.

\subsubsection{General Properties} \label{sec:genprop3Di0}

\begin{figure}
 \centering
 \includegraphics[width= \linewidth]{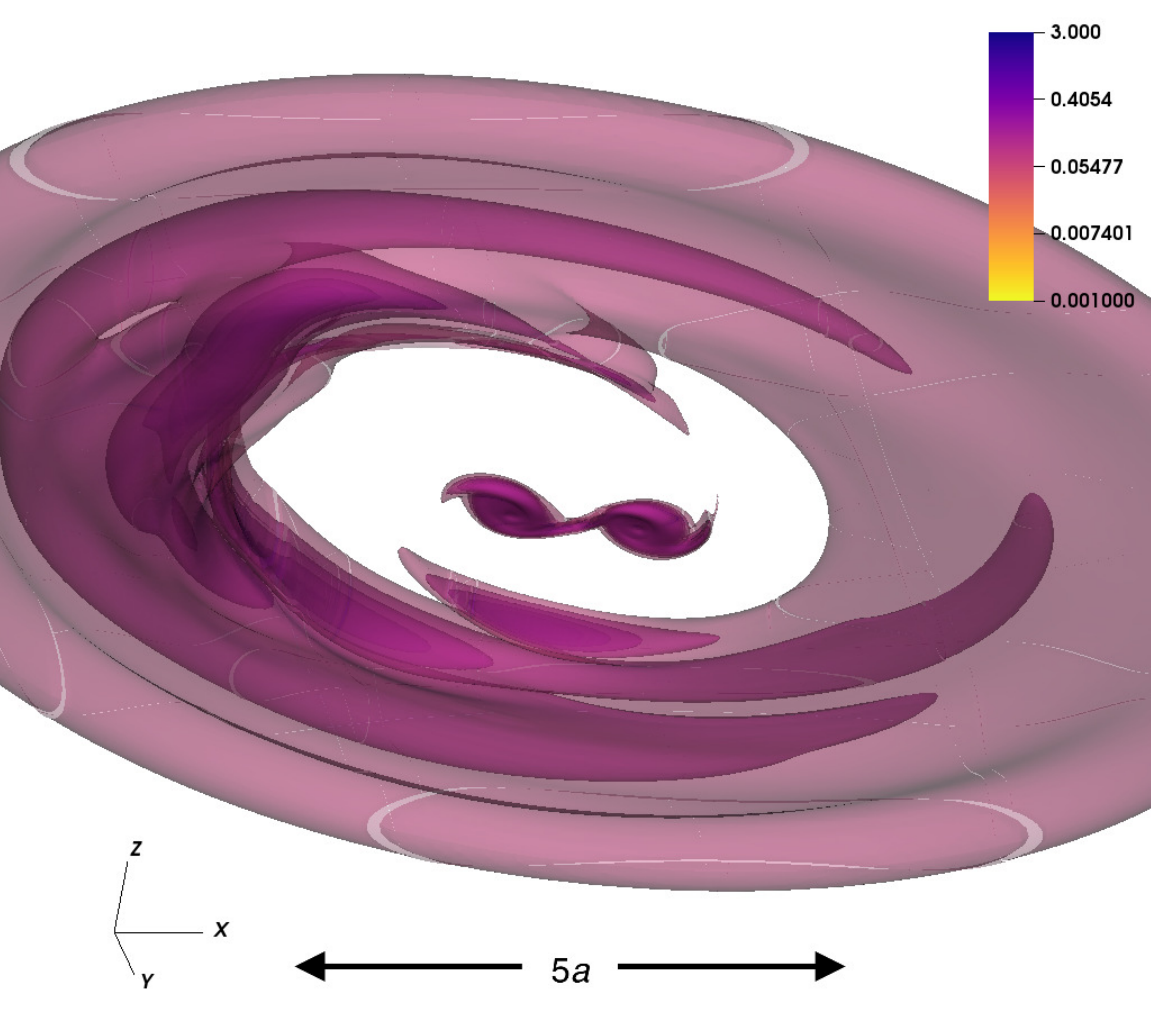}
   \caption{Snapshot of the surface density $\rho$ at $t=1000$ binary orbits for our 3D aligned simulation with resolution of $256\times256\times128$. Density is shown in a linear scale, drawn in isosurfaces.}
   \label{fig:3Di0dens}
\end{figure}

\begin{figure}
 \centering
 \includegraphics[width= \linewidth]{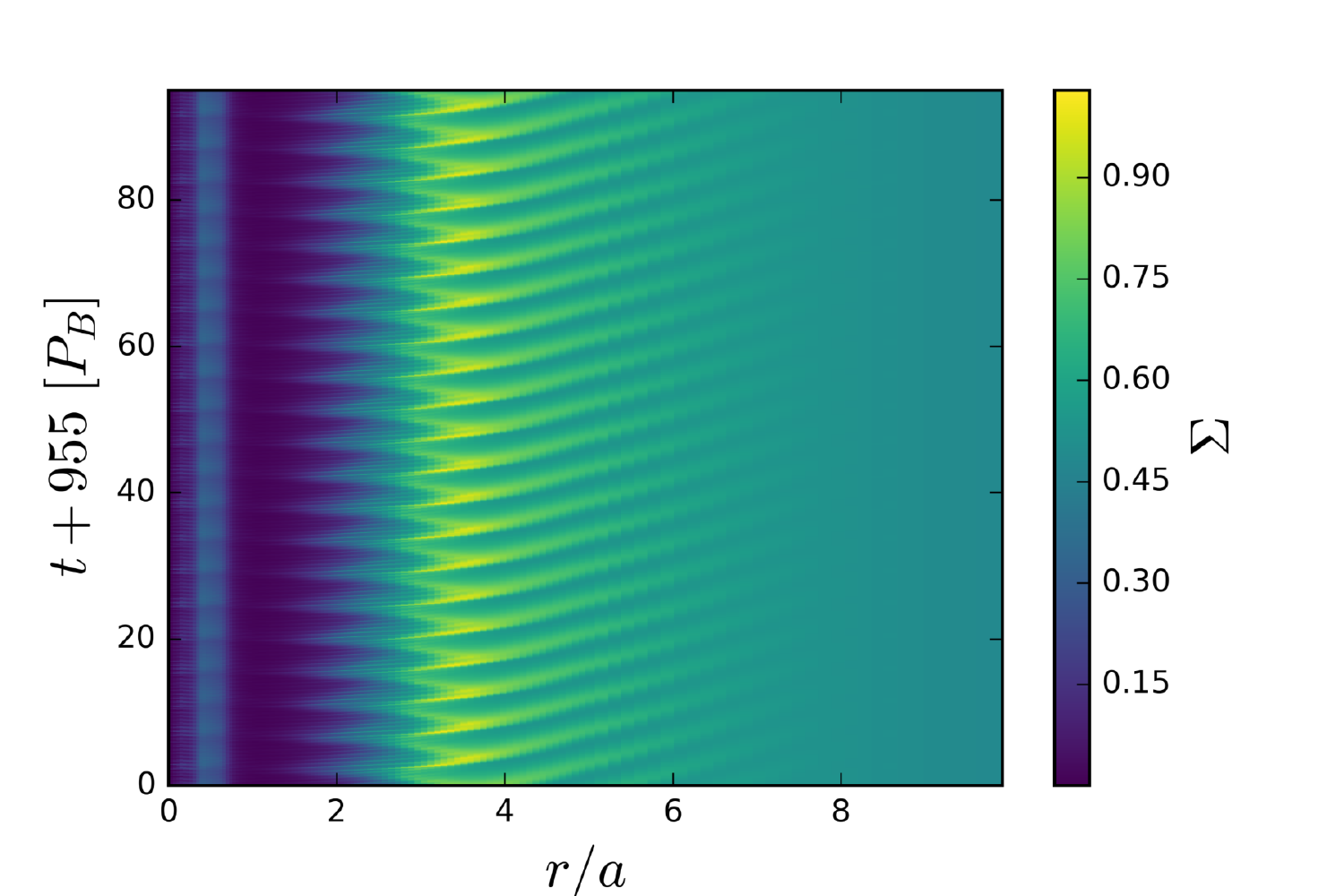}
   \caption{Spacetime diagram of surface density of the disk for the last 100 binary orbits, for the 3D aligned simulation. The $y$ axis is time in binary orbits. The regular pattern seen here indicates spiral density waves.}
   \label{fig:3Di0spacetime}
\end{figure}

Looking at the general properties of the 3D simulation, we see very similar features to the 2D disk as expected for the aligned case. Figure \ref{fig:3Di0dens} shows isosurfaces of density, with isosurface chosen to particularly illustrate the orientation of the mini disks and circumbinary disk, as well as the high density features: the mini disks, high density lump near the edge of the gap region, and the beginning part of the accretion stream. Because we use isosurfaces the gap region looks to be devoid of gas. This is simply an artifact of the isosurfaces. The structure of the disk looks very similar to the structure of the 2D disk. We examine the surface density to see if we are in a steady state. The surface density is integrated over spherical shells, and we plot a spacetime diagram of surface density in Figure \ref{fig:3Di0spacetime} similar to Figure \ref{fig:2d256spacetime} for the 2D simulation. The structure is nearly identical to the 2D case; we see spiral density waves and not much other structure. From this we expect the accretion rate and torques to behave similarly to the 2D case.

\subsubsection{Accretion Rate}

\begin{figure}
 \centering
 \includegraphics[width= \linewidth]{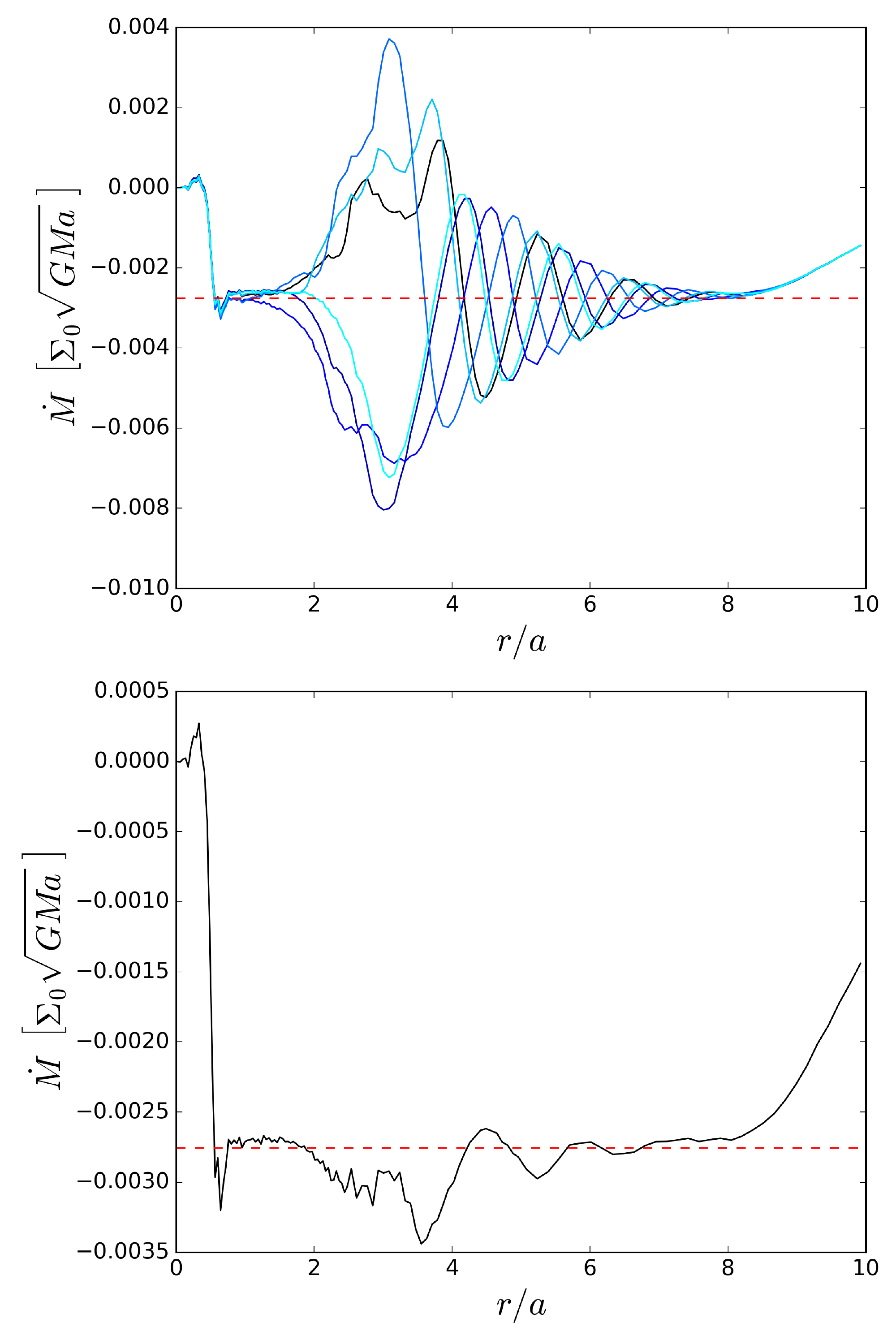}
   \caption{Accretion rate in the disk as a function of radius for the 3D aligned simulation, averaged over spherical shells. In the upper panel we plot accretion rate for the last 60 orbits, with each line representing an average over 10 orbits.  The lines follow the black to lightest blue color gradient progressively in time. The red dashed line indicates the overall average of the accretion rate, averaged over time (the last 100 orbits) and over radius (excluding the buffer zone of the disk). The lower panel shows the accretion rate through the disk, averaged over the entire period of the last 100 orbits. The red dashed line is again the overall average, averaged over radius again excluding the buffer zone. The overall average rate is 0.0026 $\Sigma_0 \sqrt{GMa}$.}
   \label{fig:3Di0mdot}
\end{figure}

The accretion rate through the 3D aligned disk is shown in Figure \ref{fig:3Di0mdot}. As in Figure \ref{fig:2d256mdot} the upper plot is the rate through the disk for the last 100 binary orbits, each line an average over 10 orbits. The lower plot is the average accretion rate over this entire period. The red dashed line in both cases is the overall average of the accretion rate, averaged over radius and time. Again we see larger oscillation on shorter timescales within the gap region and inner edge of the disk, whereas the average over the last time period (the right-hand plot of Figure \ref{fig:3Di0mdot}) is steadier. From Figure \ref{fig:3Di0spacetime} we see that again this variation is associated with the oscillation of the inner edge of the disk. The value of the accretion rate is lower than the 2D case by roughly a factor of of 5. When normalized by the density of the disk, the normalized accretion rate of the 3D aligned disk agrees with the normalized accretion rate of the 2D disk.

\begin{figure}
 \centering
 \includegraphics[width= \linewidth]{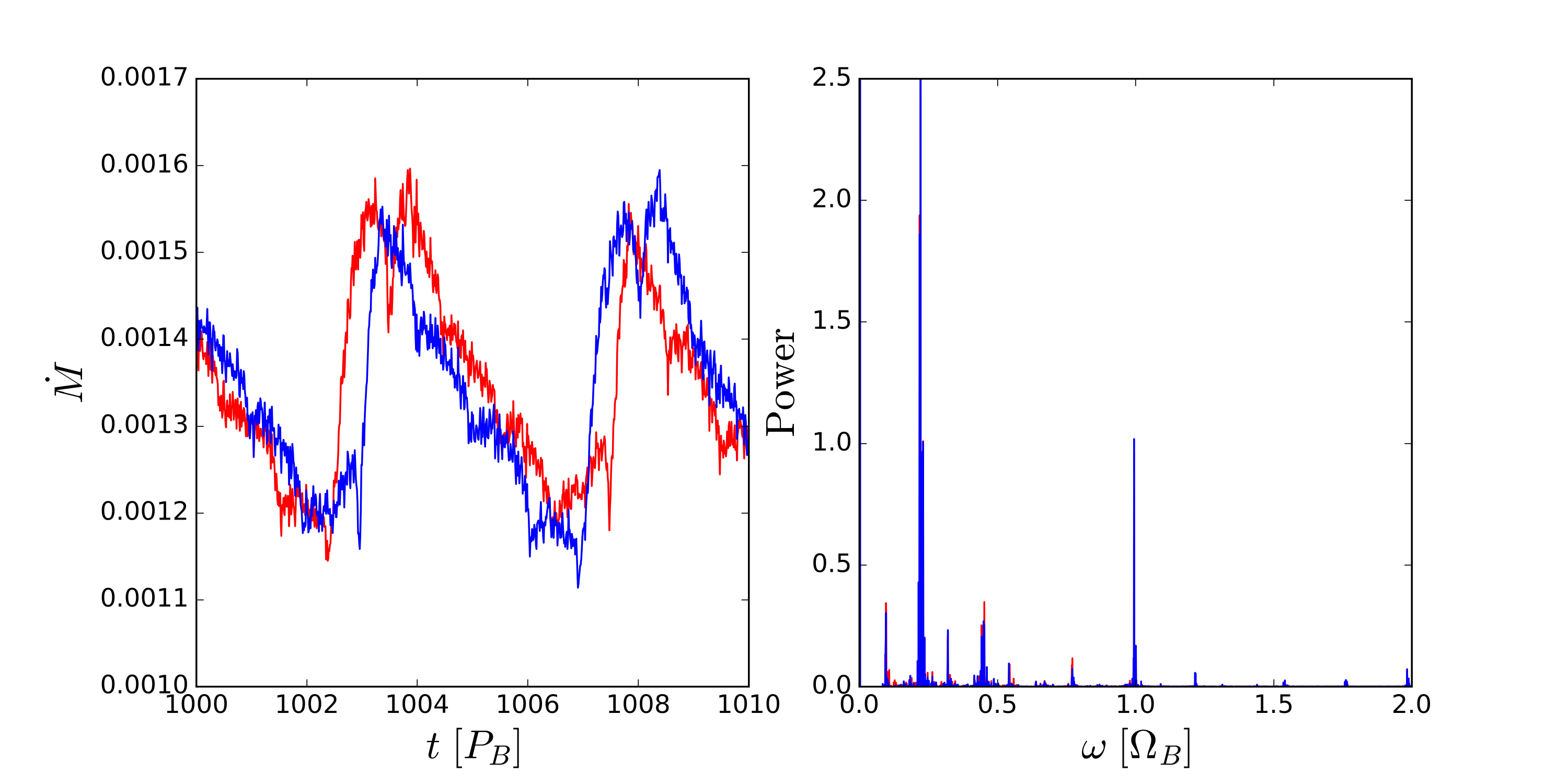}
   \caption{Accretion rate for the black holes as measured by recording the mass removed by the sink cells representing the black holes, for the 3D aligned simulation (resolution $256\times256\times128$). The left-hand panel shows the accretion rate for each of the individual black holes for a period of ten orbits near the end of the simulation. The right-hand panel shows the power spectrum of the accretion rates, for the entire simulation run time.}
   \label{fig:3Di0BHacc}
\end{figure}

As with the 2D simulation, we compare our calculation of accretion rate through the disk with the accretion rate as measured by mass removal by the sink cells representing the black holes. Each black hole has an accretion rate of 0.001195 $\Sigma_0 \sqrt{GMa}$, giving a total accretion rate of the binary of 0.00239 $\Sigma_0 \sqrt{GMa}$. This agrees with the disk accretion rate to 8\%, about the same accuracy as we found with the 2D simulation. The individual accretion rates can be seen in the left-hand panel of Figure \ref{fig:3Di0BHacc}, which shows the accretion rates over ten binary orbits. There is an overall variability on the timescale of roughly 5 or 6 binary orbits, similar to what we saw in the 2D simulation. The right-hand panel of Figure \ref{fig:3Di0BHacc} shows the power spectrum of the accretion rate for both the primary and secondary. There is a clear peak at $\omega = 0.2245$, which is roughly the same frequency fluctuation discussed in Section \ref{sec:2dmdot} and seen in MML17 ($\omega = \Omega_B/5$). Comparing the 2D (Figure \ref{fig:2D256BHacc}) and 3D simulations, we see that in the 3D simulation we do see the high frequency fluctuations, but the overall power of the 3D simulations is much lower than the 2D simulations.

\subsubsection{Torques}

\begin{figure}
 \centering
 \includegraphics[width= \linewidth]{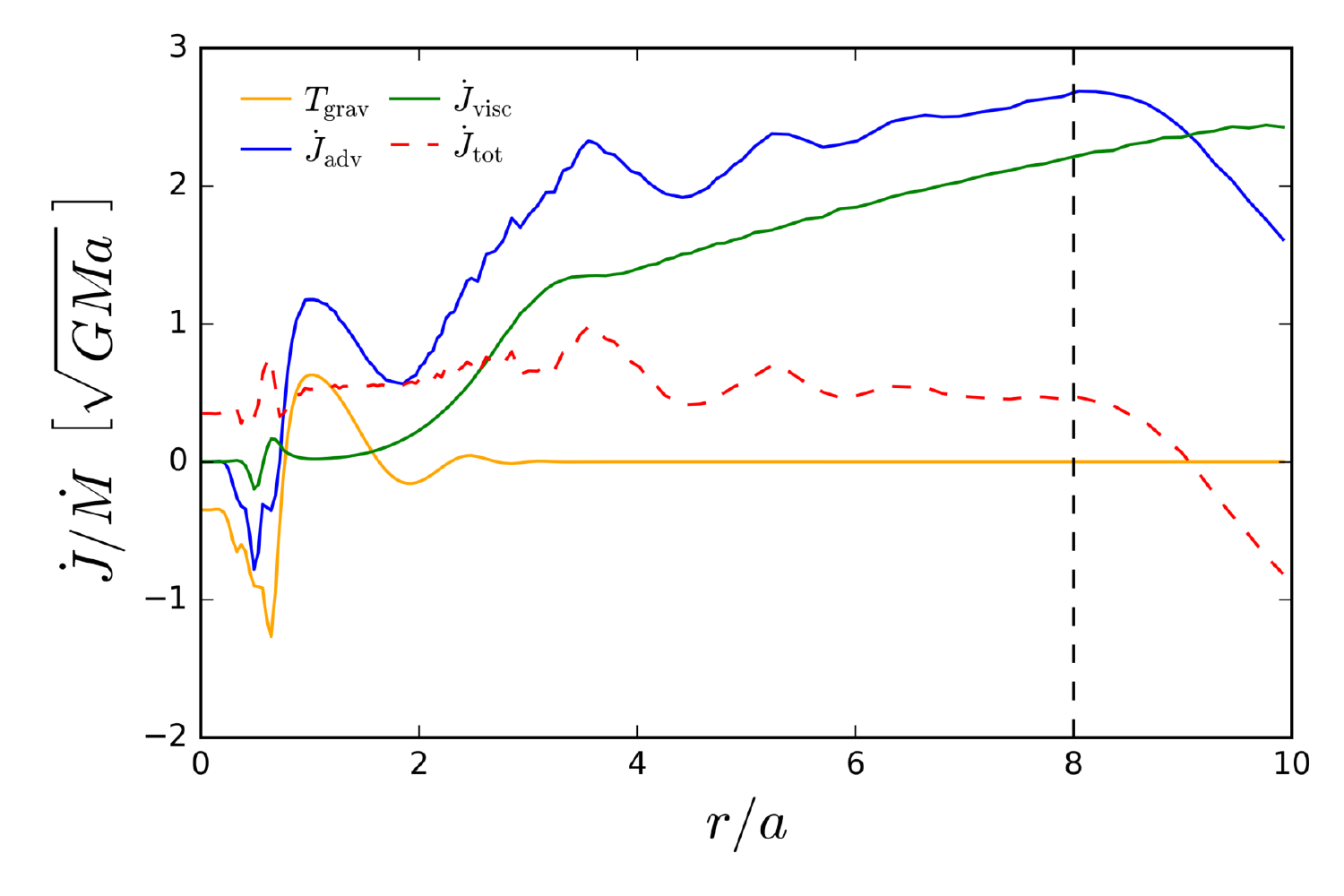}
   \caption{Advective, viscous, gravitational, and total torques on the system for the 3D aligned simulation, normalized by the average accretion rate as shown in Figure \ref{fig:3Di0mdot}.}
   \label{fig:jflux3Dinplane}
\end{figure}

Figure \ref{fig:jflux3Dinplane} shows the z-component of the torque for the aligned simulation, as we focus on the evolution of the binary which is fixed to be in the same plane as the disk midplane ($z=0$). The gravitational and viscous torque have much of the same shape and behavior that we see in the 2D simulation, shown in Figure \ref{fig:jflux2d256}. The main difference between the 2D case and the 3D aligned case lies in the advective torque, in the outer regions of the disk. There is more variation in the 3D aligned case in the region $r=4a-6a$. Additionally, Figure \ref{fig:jflux2d256} shows that the advective and viscous torques have the same slope in the region outside roughly $r=4$ for the 2D simulation, whereas Figure \ref{fig:jflux3Dinplane} shows a slightly shallower slope for the advective torque compared to the viscous torque. In this outer region, the disk is thicker which means movement in the z direction is more likely to occur than in the inner regions of the disk. This increases the x- and y-components of the advective torque, and in turn also perhaps decreasing the z-component of the torque. The disk holds equatorial symmetry, as indicated by the average z-momentum of the disk. This value varies through the simulation but is always less than 0.0006 $\Sigma_0 \sqrt{GM/a}$, which is roughly 1/1000 of the x- and y-momenta.

The total torque is still very close to constant throughout the disk, with an average value of 0.558 $\sqrt{GMa}$. This value is smaller than those found in 2D simulations, both our own and those of MML17 and \citet{munoz18}. We apply the same analysis as in Section \ref{sec:torque2D} to determine the evolution of the binary. We find a positive $\dot{a}$ in the evolution equation (equation \ref{eq:evolution}). We conclude that in 3D when the binary and the disk are in the same plane, we will see expansion of the binary over time.

%---------------------------------------
\subsection{3D Simulations - misaligned disk}
%---------------------------------------
Now that we have examined systems with the binary in the same plane as the disk, we perform simulations of misaligned systems. The disk is initialized in the $z=0$ plane, and the inclination of the binary orbital plane is changed to a fixed value. We focus on the simulation where the binary inclination is 45\degree (simulation F as described in table \ref{tab:table1}). The simulation has the same computational domain as the aligned disk case ($[-10a,10a]\times[-10a,10a]\times[-5a,5a]$) with the same resolution ($256\times256\times128$). The total run time remains the same (1050 binary orbits), and the required CPU time is 89200 hours.

\subsubsection{General Properties}

\begin{figure}
 \centering
 \includegraphics[width= \linewidth]{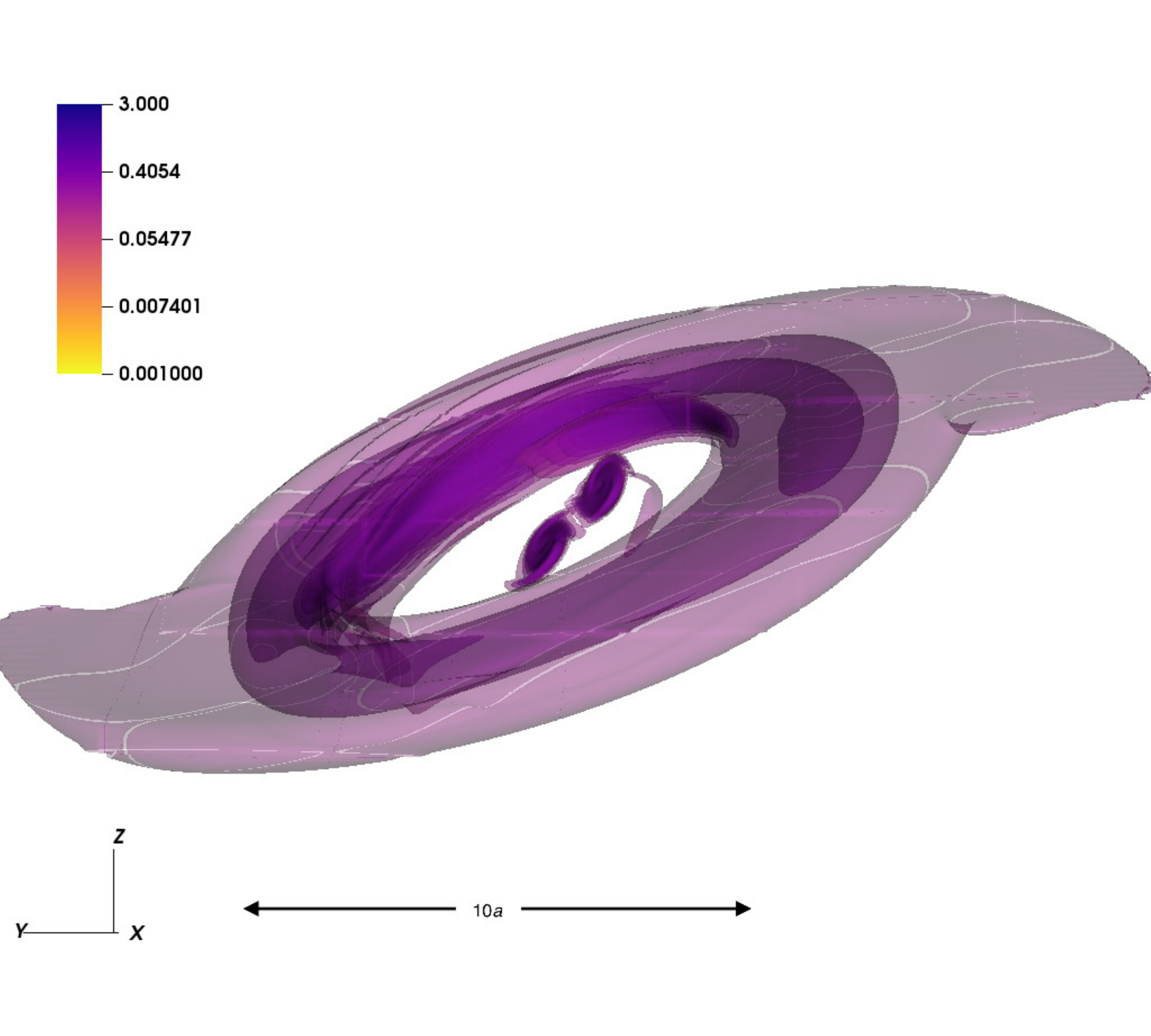}
   \caption{Snapshot of the surface density $\rho$ at $t=500$ binary orbits for our 3D misaligned simulation with resolution of $256\times256\times128$. The binary inclination is 45\degree. Density is shown in a linear scale, drawn in isosurfaces.}
   \label{fig:3Di45dens}
\end{figure}

In Figure \ref{fig:3Di45dens}, we show density halfway through the simulation to illustrate the structure of the misaligned disk. At the very beginning of the simulation, the disk is aligned in the $z=0$ plane. Very quickly the disk begins to warp, realigning itself with the orbital plane of the binary. The inner regions begin to warp first, with the warp propagating outward in the disk slowly. The very outer edges of the disk remain mostly aligned with the $z=0$ plane for much of the simulation. Mini disks again form around each individual black hole, but now the orientation of the mini disk is also variable. When the mini disks are formed they are aligned with the plane of the disk. Throughout the simulation, the disks' orientation changes to realign with the binary orbital plane, which occurs much more quickly than the realignment of the circumbinary disk. Figure \ref{fig:3Di45dens} illustrates all these properties: the outer portions of the disk are aligned with the $z=0$ plane, the mini disks are aligned with the binary orbital plane at 45\degree, and the inner portion of the disk has started to align with the binary orbital plane but has not yet reached full alignment. As stated in Section \ref{sec:genprop3Di0} the apparent lack of gas in the gap region is an artifact of the fact that we are using isosurfaces of density, which were chosen to illustrate high density features and the orientation of the circumbinary and mini disks.

\citet{nixon13} perform 3D simulations around binary black holes, using a smoothed particle hydrodynamics code. They study a range of inclinations for equal mass binaries on circular orbits. In their simulations they see tearing of the disk for misaligned disks. The inner region of our disks show some disruption, though not as dramatic tearing as what \citet{nixon13} see. To estimate the breaking radius of the disk, i.e. where the disk will tear. \citet{nixon13} give the condition that the disk will break when the viscous torque in the disk is less than the precession torque. Using the analytic expressions for both (they do not calculate torques within their simulations), they give the disk breaking radius with the following expression:
\begin{equation}
    R_{\mathrm{break}} \lesssim 50 \mu^{1/2} {|\sin{2\theta}|}^{1/2} {\Big(\frac{H/R}{10^{-3}}\Big)}^{-1/2} {\Big(\frac{\alpha}{0.1}\Big)}^{-1/2} a.
\end{equation}
They use a disk aspect ratio of $h=0.01$, which indicates that their entire disk (which extends to $r=8a$) will break. Our simulation uses an aspect ratio of $h=0.1$ which gives a breaking radius of about $3.5a$. Within this radius we primarily have the gap region and the very inner edge of the disk, so most of our disk will remain intact and not tear, hence why we see much less dramatic tearing than \citet{nixon13}.

\subsubsection{Accretion Rate}

\begin{figure}
 \centering
 \includegraphics[width= \linewidth]{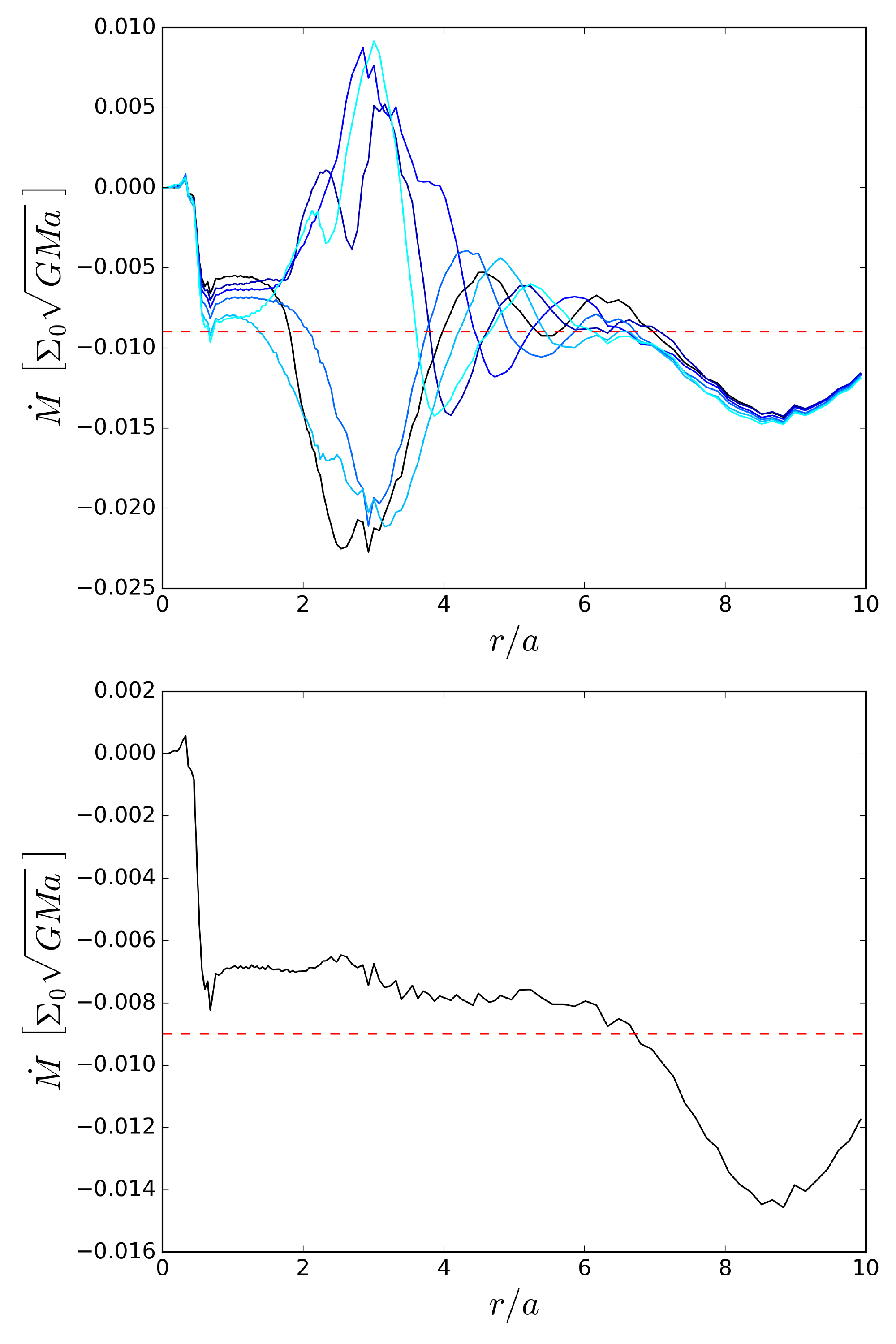}
   \caption{Accretion rate in the disk as a function of radius for the 3D 45\degree inclined simulation run. The accretion rate is averaged over a spherical shell. In the upper panel, we plot accretion rate for the last 60 orbits, with each line representing an average over 10 orbits.  The lines follow the black to lightest blue color gradient progressively in time. The red dashed line indicates the overall average of the accretion rate, averaged over time (the last 100 orbits) and over radius (excluding the buffer zone of the disk). The lower panel shows the accretion rate through the disk, averaged over the entire period of the last 100 orbits. The red dashed line is again the overall average, averaged over radius again excluding the buffer zone. The overall average rate is 0.00899 $\Sigma_0 \sqrt{GMa}$.}
   \label{fig:3di45mdot}
\end{figure}

\begin{figure}
 \centering
 \includegraphics[width= \linewidth]{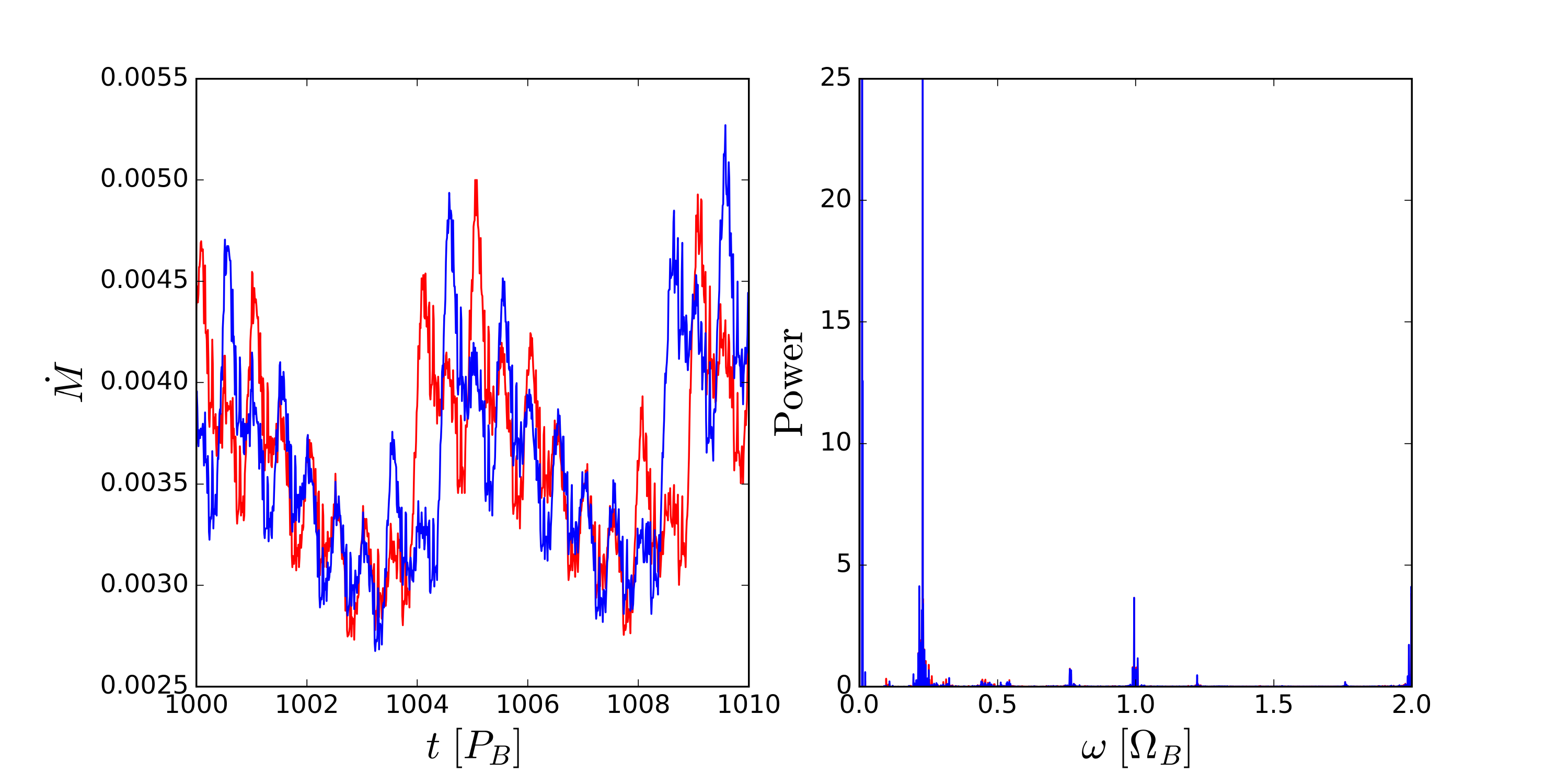}
   \caption{Accretion rate for the black holes as measured by recording the mass removed by the sink cells representing the black holes, for the 3D misaligned simulation (resolution $256\times256\times128$). The left-hand panel shows the accretion rate for each of the individual black holes for a period of ten orbits near the end of the simulation. The right-hand panel shows the power spectrum of the accretion rates, for the entire simulation run time.}
   \label{fig:3Di45BHacc}
\end{figure}

The accretion rate through the disk is shown in Figure \ref{fig:3di45mdot}. In the upper panel we have the accretion rate plotted in averages over 10 orbits for the last 100 orbits of the simulation, and in the lower panel we have the average accretion rate over that entire 100 orbit period. We see similar behavior in this misaligned case that we saw previously in the aligned case. One difference is the value of the average accretion rate throughout the disk, 0.00899 $\Sigma_0 \sqrt{GMa}$, is larger than the 3D aligned simulation, though the two rates are still within an order of magnitude. \citet{nixon13} find that their misaligned disks have much higher accretion rates compared to the aligned disk, i.e. tearing of the disk enhances the accretion rate. Their 45\degree inclined simulation has an accretion rate roughly 50 times the aligned simulation accretion rate. While our simulations also show an increased accretion rate in the 45\degree case, it is slightly less than a factor of four larger than the aligned case. 

Figure \ref{fig:3Di45BHacc} shows the accretion rate as measured by the mass removal of the sink cells. Both plots look very similar to the 3D aligned case - the frequency spectrum peaks at $\omega = 0.229$ indicating the existence of a density lump in this misaligned case. The misaligned case shows more variation at higher frequencies than the aligned case.

\subsubsection{Torques}

\begin{figure}
 \centering
 \includegraphics[width= \linewidth]{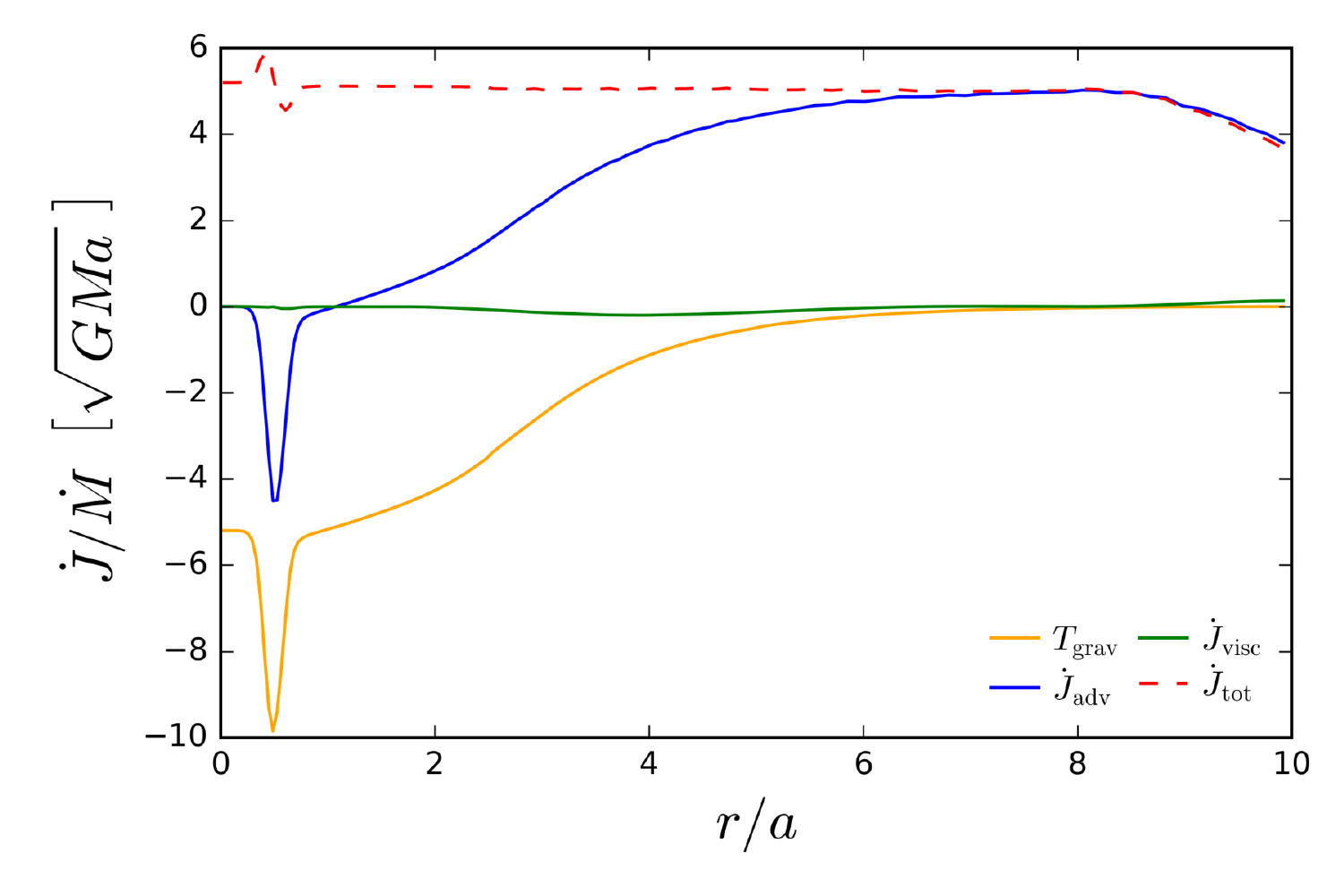}
   \caption{Advective, viscous, gravitational, and total torques on the system for the 3D simulation with binary and disk planes misaligned by 45\degree, normalized by the average accretion rate. The normalized torques are the z-components of torque in the reference frame of the binary orbital plane.}
   \label{fig:jfluxbin45}
\end{figure}

In order to examine the torques for the misaligned system, we calculate the torque components in the same way as done previously, and then project the components into the binary orbital frame. We then take the z-components of the torques in the binary frame, and perform the same analysis to determine the binary evolution. The z-components in the binary frame are shown in Figure \ref{fig:jfluxbin45}, along with the total torque. We can see that the shapes of the components have vastly different form compared to the in-plane case. As our disk is misaligned, the torque is no longer dominated by the z-component only.

We see from Figure \ref{fig:jfluxbin45} that the total torque is remarkably constant throughout the disk. Applying the analysis described in Section \ref{sec:torquecalc}, we calculate an $l_0$ value of 4.883, which is greater than $3l_B/8$, again indicating expansion of the binary orbit. We conclude that in all cases, a viscous disk will cause the binary orbit to expand.

\subsubsection{Disk Realignment}

\begin{figure}
 \centering
 \includegraphics[width= \linewidth]{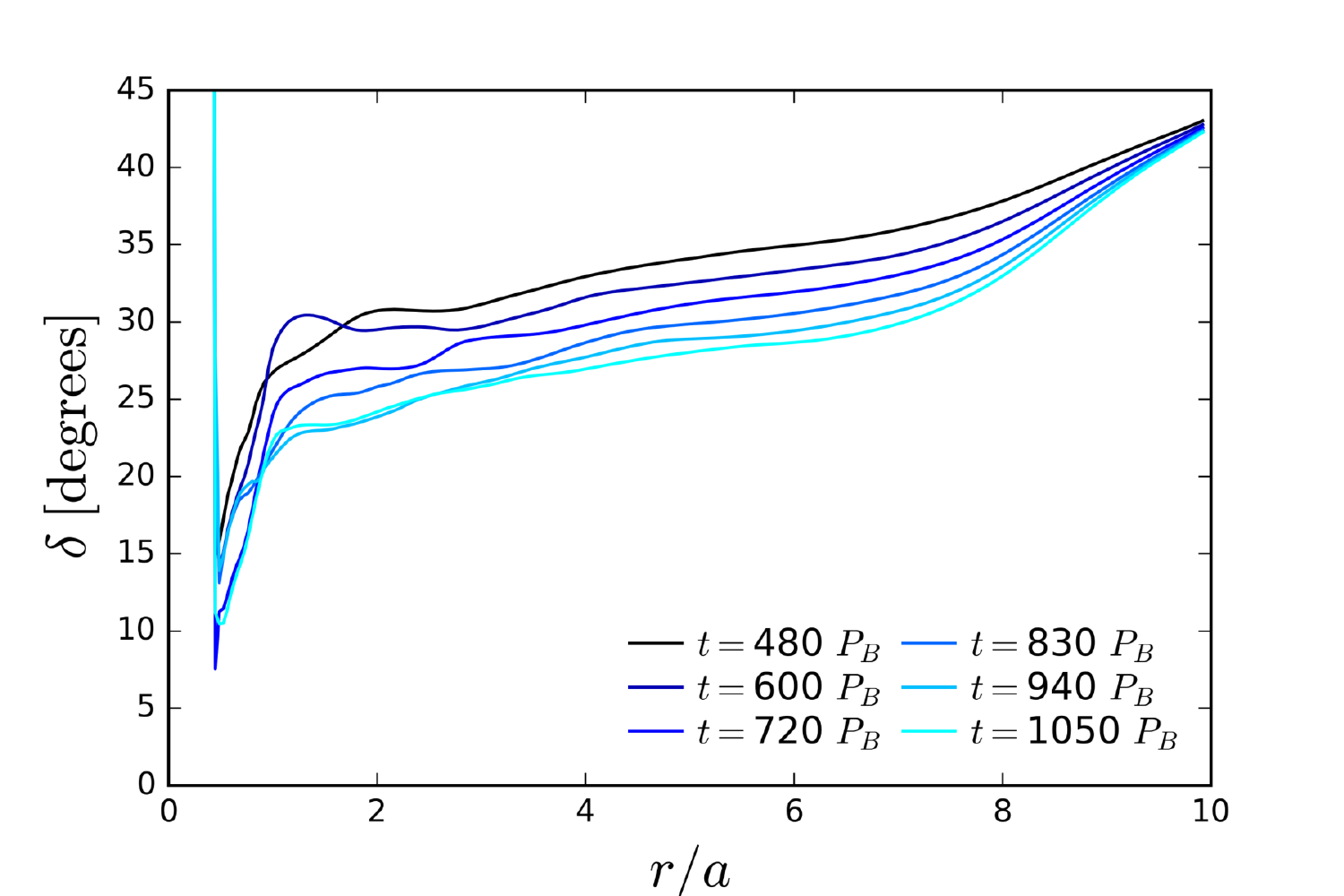}
   \caption{Inclination angle of the disk relative to the binary orbital plane as a function of radius, for the 45\degree inclined simulation. The various curves show the inclination angle at different times through the second half of the simulation.}
   \label{fig:incvsrad}
\end{figure}

We expect the disk to realign with the plane of the binary. In Figure \ref{fig:incvsrad} we show the inclination angle of the disk as a function of radius, where the angle is the relative angle between the disk and the binary orbital plane. The different curves show inclination as a function of radius for different times in the second half of the simulation. The disk realignment is clearly seen from the movement of these different curves towards lower relative inclination angle. We can also see that the inner portion of the disk is realigning more quickly than the outer portion of the disk. One significant feature is that the buffer zone seems to be acting as an anchor preventing the very outer region of the disk from realigning. This could potentially be affecting the inner portions of the disk, and future work should include a much larger disk or different treatment of the buffer zone to limit this effect.

\begin{figure}
 \centering
 \includegraphics[width= \linewidth]{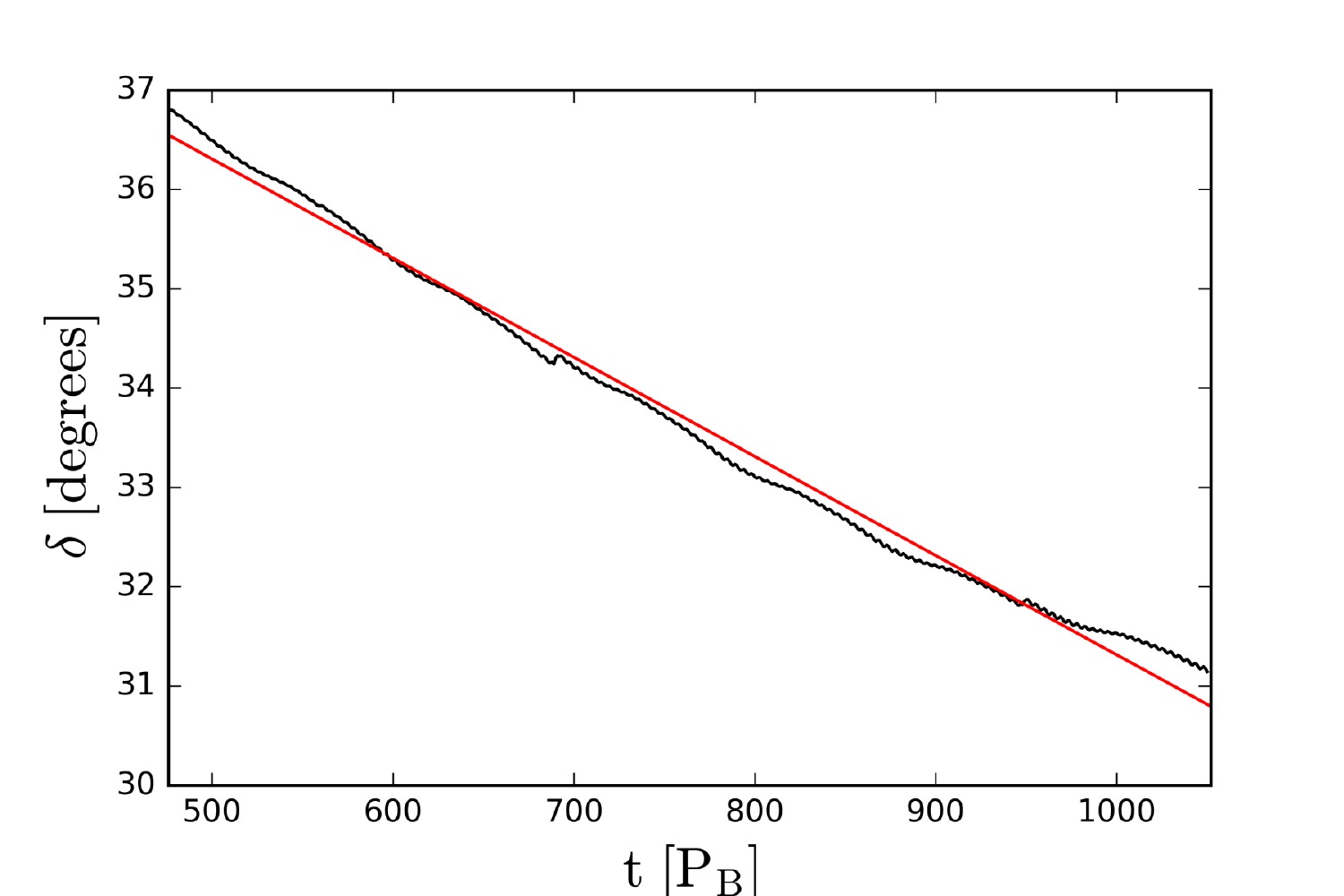}
   \caption{Inclination angle of the disk relative to the binary orbital plane as a function of time, for the 45\degree inclined simulation. The black line is the simulation data and the red line is the least squares fit to the data, to calculate realignment time of the disk.}
   \label{fig:realign45}
\end{figure}

In Figure \ref{fig:realign45} we show the inclination angle of the disk relative to the binary orbital plane over time for half of the total simulation time but only in the outer region of the disk. This region is from $r=7a$ to $r=8a$ (before the buffer zone begins) in order for us to examine the global evolution of the disk. We perform a least squares regression fit to this curve to calculate the realignment time, and find a time of $\sim$ 4134 binary orbits. The global viscous time of the disk (calculated for $r=10a$) is $\sim$ 5020 binary orbits, so $t_{realign}/t_{visc} \approx 0.8236$. If we compare the realignment time to the viscous time at $r=7.5a$, we find $t_{realign}/t_{visc7.5} \approx 1.271$. This result is in good agreement with the work of \citet{fragner10}, which finds the realignment time to be roughly on the timescale of the viscous evolution. \citet{foucart14} study the evolution of warp in accretion disks analytically, and apply their findings to circumbinary disks. They find that circumbinary disks should realign on the global precession timescale, which is much shorter than the viscous timescale. Their surface density profile that is steeper than we use, and they note that shallower profiles make it more difficult for torques applied on the smaller amount of material at small radii to affect the global evolution of the disk. Shallower surface density profiles therefore will have longer realignment timescales (than the precession time) which is consistent with our results.

In the inner regions of the disk we perform the same analysis, and find as expected the inner regions realign more quickly than the outer regions, but are still within the same magnitude of the viscous evolution time.

%==================================================
%==================================================
\section{Summary and Discussion}
%==================================================
%==================================================

%---------------------------------------
\subsection{Summary}
%---------------------------------------
We have performed hydrodynamic simulations of circumbinary accretion disks using the grid-based finite volume code \textsc{athena++}. The binary is on a fixed circular orbit with an equal mass ratio, and individual members of the binary are treated as sink cells. The disks are locally isothermal and use an $\alpha$ viscosity. We performed 2D simulations, and 3D simulations for both aligned and misaligned disks and study the mass accretion rate and the torques on the system.

The mass accretion rate in all three simulations shows steady accretion with variation on short timescales which is associated with a high density lump on the inner edge of the disk. The 2D and 3D aligned disk simulations have comparable accretion rates though the 3D rate is roughly a factor of 5 smaller than the 2D rate. This is due to the density profiles through the disk. As mentioned in Section \ref{sec:numsetup} the surface density of both disks has the same scaling but the full density profile scales differently, resulting in slightly different overall surface density profiles and therefore slightly different values of the accretion rate. The 3D misaligned disk shows an enhanced accretion rate compared to the aligned disk, possibly due to some tearing of the disk.

The net torque on the system indicates in all simulations that the binary orbit will expand. This confirms some previous results in 2D. The different components of torque show the same behavior in 2D and the 3D aligned disk. The 3D misaligned disk shows very different behavior in the individual components of the torque, but as stated the net torque is still positive and indicates expansion of the binary.

The overall sign of the net torque is very sensitive to the value of the individual torque components, in particular the advective torque. This indicates that a careful examination of the value of $\alpha$ is needed. As mentioned briefly in Section \ref{sec:numsetup} MML17 carry out simulations of circular binaries using $\alpha = 0.1$ and $\alpha = 0.05$ and find the same result for both values of $\alpha$. In recent MHD simulations of accretion disks it is found that the equivalent $\alpha$ that results from turbulence in such disks is around 0.1 so this value is reasonable \citep{zhu18}.

%---------------------------------------
\subsection{Implications for Binary and Disk Evolution}
%---------------------------------------
It is a common assumption that gaseous disks are the mechanism by which supermassive black hole binaries are able to shrink their semi-major axes to less than a parsec \citep{begelman80, escala05}. As discussed in Section \ref{sec:torquecalc}, we analyze the torques in the simulation to determine if the binary will expand or shrink. In all of our simulations, both 2D and 3D, we find that the torques on the system cause the binary to grow, contrary to some previous results. This suggest that hydrodynamic circumbinary disks cannot be a solution to the final parsec problem for supermassive black hole binaries, at least for equal mass binaries. Perhaps different configurations of the disk may provide a solution, such as different mass ratios, eccentricities, or more massive disks. For equal mass binaries however, it seems that circumbinary accretion disks will prevent the merger of these binaries.

To examine this more closely we compare the rate of change of the semimajor axis due to the gas disk (which we find widens the binary) with the rate of change due to dynamical friction from stars (which shrinks the binary). To calculate the rate of change from the gas disk we use equation \ref{eq:evolution} using typical values for a binary system. We use a mass of $10^6 M_{\odot}$ for each black hole and a binary separation ($a$) of one parsec. The $\dot{M}$ and $l_0$ values are from our 3D aligned simulation (simulation E in Table \ref{tab:table1}), which are 0.0026 $\Sigma_0 \sqrt{GMa}$ and 0.558 $\sqrt{GMa}$ respectively. To calculate the rate of change due to stars we use equation 8.20 from \citet{bandt}:
\begin{equation}
\frac{d}{dt} \Big(\frac{1}{a}\Big)_{\textrm{stars}} = -14.3 \frac{G\rho}{\sigma}
\end{equation}
or the following which is in a more directly comparable form to equation \ref{eq:evolution}
\begin{equation}
\Big(\frac{da}{dt}\Big)_{\textrm{stars}} = 14.3 \frac{G\rho}{\sigma} a^2 \ .
\end{equation}
$\rho$ is the density of stars and $\sigma$ is the velocity distribution of the stars. We choose a typical stellar density of $3.386 \times 10^{-24}$ g/cm$^3$ which corresponds to 0.1 stars per cubic parsec and average stellar mass of 0.5$M_{\odot}$.The velocity dispersion is set to 75 km/s. Using these parameters, we find that the two rates are comparable if the mass of the disk is around 0.04 times the combined mass of the black holes. For the case where the mass of the disk is 0.1 the mass of the black holes, $({da}/{dt})_{\textrm{gas}} = 11.24$ cm/s and $({da}/{dt})_{\textrm{stars}} = 4.100$ cm/s. This is significant as it indicates that the circumbinary disk can potentially slow or stall the shrinking of the binary.

We found the length of the disk realignment time to be on the same timescale as the global viscous time of the disk, in good agreement with the results of \citet{fragner10}. Their disk configuration has the primary at the center of the disk, with the secondary at a distance of $30a$ from the primary. This leads us to conclude that the alignment evolution of the disk in any configuration will be on the viscous evolution time.

%---------------------------------------
\subsection{Limitations and Future Work}
%---------------------------------------
One major limitation of this work is the absence of magnetic fields. We use the $\alpha$-disk prescription for viscosity, and in reality we expect the disk viscosity to be driven by turbulence, resulting from the magnetorotational instability \citep{balbus91}. Some magnetohydrodynamic simulations have already been performed, by \citet{shi12}. They perform 3D simulations of an aligned disk around an equal mass binary on a circular orbit, and find that the binary shrinks. However, they cut out the central region of the simulation which includes the binary. In future work, magnetic fields should be included and full 3D simulations for a range in binary-disk inclinations explored. Given that various results indicate the importance of the advective torque to the evolution of the binary, it is even more important that MHD calculations be performed.

In addition, several assumptions made should be relaxed or explored further in the future. This includes the binary mass ratio, disk aspect ratio, and disk mass. We also use a local isothermal equation of state, so in future work this should change to include a more detailed treatment of the thermodynamics to include the cooling time of the gas. Though we assume the binary orbit to be fixed as the evolution of the binary does not change the eccentricity significantly, a careful treatment would include coupling the evolution of the binary with the evolution of the disk, so the semi-major axis and eccentricity are allowed to change as the simulation runs. Additionally, we do not include the self-gravity of the disk or general relativistic effects, which could significantly affect the dynamics of the binary-disk interaction.

%% Included in this acknowledgments section are examples of the
%% AASTeX hypertext markup commands. Use \url without the optional [HREF]
%% argument when you want to print the url directly in the text. Otherwise,
%% use either \url or \anchor, with the HREF as the first argument and the
%% text to be printed in the second.

%==================================================
%==================================================
\acknowledgments
We thank Ryan Miranda, Diego Mu\~{n}oz, Dong Lai, Zolt\'{a}n Haiman, Alwin Mao, Goni Halevi, and Cole Holcomb for helpful input and discussions. We also thank the referee for helpful comments that have improved our paper.

\bibliographystyle{apj}
\bibliography{biblio}
\clearpage

\clearpage

%% Any table notes must follow the \end{tabular} command.

%%% change \tablenotetext{a} -> \tablenotetext{1} etc. to omit LaTeX errors
%%% Karol Kozioł for ShareLaTeX

%% If the table is more than one page long, the width of the table can vary
%% from page to page when the default \tablewidth is used, as below.  The
%% individual table widths for each page will be written to the log file; a
%% maximum tablewidth for the table can be computed from these values.
%% The \tablewidth argument can then be reset and the file reprocessed, so
%% that the table is of uniform width throughout. Try getting the widths
%% from the log file and changing the \tablewidth parameter to see how
%% adjusting this value affects table formatting.

%% The \dataset macro has also been applied to a few of the objects to
%% show how many observations can be tagged in a table.

\clearpage

%% Tables may also be prepared as separate files. See the accompanying
%% sample file table.tex for an example of an external table file.
%% To include an external file in your main document, use the \input
%% command. Uncomment the line below to include table.tex in this
%% sample file. (Note that you will need to comment out the \documentclass,
%% \begin{document}, and \end{document} commands from table.tex if you want
%% to include it in this document.)


%% The following command ends your manuscript. LaTeX will ignore any text

\end{document}